\documentclass[aps,prd,nofootinbib,reprint,superscriptaddress,preprintnumbers]{revtex4-2}
\pdfoutput=1
\usepackage[utf8]{inputenc}
\usepackage{amsmath,amssymb}
\usepackage{epsfig}
\usepackage{comment}
\usepackage{graphicx}
\usepackage[colorlinks,citecolor=blue]{hyperref}
\usepackage{xcolor}
\usepackage{subfigure}
\usepackage{cleveref}
\usepackage{url}
\usepackage{dcolumn}
\usepackage{bm}
\usepackage{soul}
\usepackage{slashed}
\usepackage{siunitx}
\DeclareSIUnit\barn{b}
\DeclareSIUnit\electronvolt{eV}
\usepackage[normalem]{ulem}

\newcommand{\dd}[1][]{\mathrm{d}^#1}

\begin{document}

\preprint{FERMILAB-PUB-24-0318-T}

\title{Tau Tridents at Accelerator Neutrino Facilities}

\author{Innes Bigaran}
\email{ibigaran@fnal.gov}
 \affiliation{Department of Physics \& Astronomy, Northwestern University, 2145 Sheridan Road, Evanston, IL 60208, USA}
\affiliation{Theoretical Physics Department, Fermilab, P.O. Box 500, Batavia, IL 60510, USA}

\author{P. S. Bhupal Dev}
\email{bdev@wustl.edu}
\affiliation{Department of Physics and McDonnell Center for the Space Sciences, Washington University, St. Louis, MO 63130, USA}

\author{Diego Lopez Gutierrez}
\email{d.lopezgutierrez@wustl.edu}
\affiliation{Department of Physics and McDonnell Center for the Space Sciences, Washington University, St. Louis, MO 63130, USA}

\author{Pedro A. N. Machado}
\email{pmachado@fnal.gov}
\affiliation{Theoretical Physics Department, Fermilab, P.O. Box 500, Batavia, IL 60510, USA}

\begin{abstract}
We present the first detailed study of Standard Model neutrino tridents involving tau leptons at the near detectors of accelerator neutrino facilities. 
The rates of these processes were previously thought to be negligible, even at future facilities. 
Our full $2\to 4$ calculation, including both coherent and incoherent scatterings, reveals that 
the DUNE near detector could observe a considerable number of tau tridents -- an important background to new physics searches. 
We identify promising kinematic features that may allow distinction of tau tridents from the usual neutrino charged-current background at DUNE, and thus establish the observation of tau tridents for the first time.
We also comment on the detection prospects at other accelerator and collider neutrino experiments. 
\end{abstract}

\maketitle
\section{Introduction}
 The origin of neutrino flavor-oscillations, and thus neutrino masses, remains one of the greatest unsolved problems with the Standard Model~(SM) of particle physics. In order to reconcile the empirical evidence of nonzero neutrino masses with the SM, a detailed study of the physics of the neutrino sector is imperative. A drawback of neutrino interaction studies is the weak interaction strengths of neutrinos with other SM particles, necessitating experimental programs with extraordinarily large statistics. The next-generation accelerator-neutrino experiments at the Intensity Frontier, such as DUNE~\cite{DUNE:2015lol} and T2HK~\cite{Hyper-Kamiokande:2018ofw}, are poised to shed light on the neutrino sector in unprecedented ways, unveiling further information about the interactions of SM neutrinos and perhaps yielding a portal to physics beyond the SM~(BSM).

Of all SM particle species, the tau neutrino~($\nu_\tau$) remains the least well-studied. 
The low cross sections for interactions, high tau-production thresholds, and experimental difficulty of distinguishing it from other neutrino states make studies of $\nu_\tau$ extraordinarily difficult. 
To date, the global dataset of directly observed $\nu_\tau$ interactions consists of only $\mathcal{O}(10)$ events at  the DONuT~\cite{DONuT:2007bsg} and OPERA~\cite{OPERA:2018nar} experiments, making future $\nu_\tau$ observation highly significant and a key target for accelerator-neutrino programs.

\begin{figure}[t!]
    \centering
\includegraphics[width=0.45\textwidth]{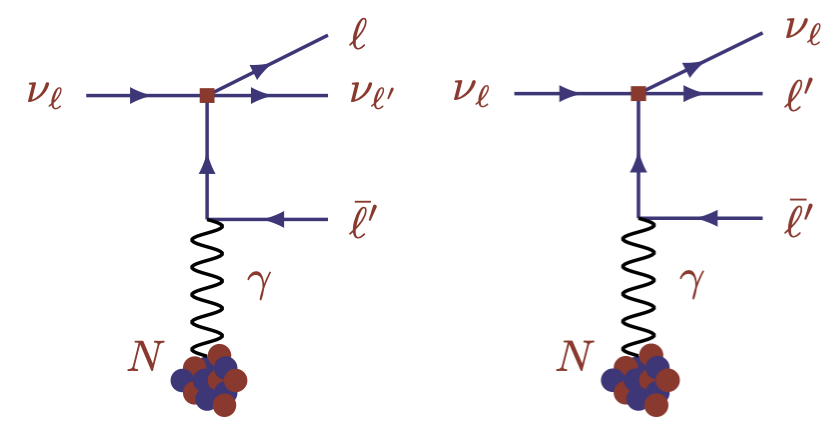}
    \caption{Neutrino trident production via charged- (left) and neutral-current (right) four-fermion interactions in the SM.}
    \label{fig:NTP}
\end{figure}

Even more challenging than simply observing $\nu_\tau$ charged-current (CC) interactions is to study neutrino trident production involving taus.
Neutrino tridents are SM processes involving the creation of a charged-lepton pair via an energetic neutrino scattering with a nucleus, $\nu_\ell N \to \nu_{\ell'} \ell \ell' N $~\cite{Czyz:1964zz} (see Fig.~\ref{fig:NTP}). 
Although rare, these processes provide a precision test of the SM weak sector, as they involve both $W$ and $Z$-mediated interactions (see e.g.,~\cite{Brown:1972vne, CHARM-II:1990dvf, CCFR:1991lpl, NuTeV:1999wlw, deGouvea:2019wav}). They are also relevant for BSM physics searches, and in particular, tridents with $\ell\neq \ell'$ will be an important background for charged lepton-flavor violation searches (see e.g.,~\cite{Altmannshofer:2014pba, Magill:2017mps,Ballett:2019xoj, Shimomura:2020tmg, Cheng:2022jyi, Bigaran:2022giz}).  
So far, only the dimuon trident $\nu_\mu N \to \nu_\mu \mu^+\mu^- N $ has been observed at $\sim 3\sigma$ level by CHARM~\cite{CHARM-II:1990dvf} and CCFR~\cite{CCFR:1991lpl} experiments (see also NuTeV~\cite{NuTeV:1998khj, NuTeV:1999wlw}), whereas the other $\ell\ell'$ final states are yet to be observed. 
This has motivated new proposals for observing tridents using future accelerator~\cite{Magill:2016hgc, Ballett:2018uuc, Altmannshofer:2019zhy} and collider~\cite{Francener:2024wul, Altmannshofer:2024hqd} neutrino experiments, as well as at neutrino telescopes~\cite{Ge:2017poy, Zhou:2019frk}.
Typically in the literature, studies have focused on the electron and muon tridents due to their relatively smaller energy threshold. 
The few existing studies~\cite{Formaggio:2012cpf,Magill:2016hgc,Zhou:2019vxt} which briefly mentioned tau tridents have highlighted their low likelihood in standard accelerator-neutrino setups originating from $\nu_\mu$ scattering, particularly at near detectors. 

A key challenge for both $\nu_\tau$ CC interactions and tau tridents is the reconstruction of the tau lepton.
One way to achieve this is by having excellent spatial resolution, at the micrometer level or below, which is typically attained using emulsion detectors, such as those at FASER$\nu$~\cite{FASER:2020gpr}.
This allows observation of the vertex displacement from tau propagation.
Alternatively, detailed event reconstruction 
capabilities, as is the case for Liquid Argon Time Projection Chambers (LArTPCs) such as in DUNE~\cite{DUNE:2015lol}, 
can lead to excellent signal-to-background ratios even if the displaced vertex from tau decay  cannot be resolved~\cite{Machado:2020yxl, Isaacson:2023gwp}. 

In this work, we perform the first detailed study of neutrino tridents involving one or two tau leptons at 
neutrino accelerator facilities, focusing specifically on the DUNE Near Detector (ND)~\cite{DUNE:2021tad}, at which the tau tridents were neglected in previous studies. 
We consider the DUNE standard CP-optimized flux, as well as the tau-optimized flux~\cite{Fields}, for our calculation. We find non-negligible number of tau tridents at DUNE-ND in both cases, and especially so in the proposed tau-optimized mode~\cite{DUNE:2020lwj}.  We also highlight the different behavior of the tau-containing trident cross sections due to a non-negligible final-state charged-lepton mass, in contrast with trident processes with lighter leptons better studied for accelerator facilities in the literature. The prospective detection of $\nu_\mu N \to \nu_{\tau} \mu \tau N $ events is promising due to low intrinsic tau neutrino background and significantly different kinematic distributions between the tau trident signal and $\nu_\mu$ CC  backgrounds. \\

The rest of the paper is organized as follows: In Section~\ref{sec:xsec}, we calculate the trident production cross sections. In Section~\ref{sec:III}, we estimate the number of trident events at DUNE. In Section~\ref{sec:ID}, we discuss how tau tridents can be identified at DUNE. Our conclusions are given in Section~\ref{sec:con}. In Appendix~\ref{app:epa}, we remark on the Equivalent Photon Approximation (EPA) approach to trident calculations. In Appendix~\ref{app:ff}, we outline the nuclear and nucleon form factors. In Appendix~\ref{app:flux}, we show the neutrino fluxes used in our calculation. In Appendix~\ref{app:exp}, we give the number of trident events at other accelerator/collider neutrino experiments. Finally, Appendix~\ref{app:dist} shows the tau trident event distributions at DUNE.

\section{Neutrino trident production} \label{sec:xsec}
Neutrino tridents are exceptionally rare $2\to 4$ electroweak scattering processes in the SM~\cite{Czyz:1964zz}. 
There are two key regimes to consider for tridents: the coherent scattering of neutrinos off nuclei without resolving their substructure, and the incoherent or diffractive scattering of neutrinos off individual nucleons. 
Coherent scattering cross sections are expected to dominate so long as the momentum transfer $Q$ is significantly lower than the inverse-nuclear radius~\cite{Czyz:1964zz}. 
For larger $Q$, incoherent interactions become relevant where the leptons interact with individual nucleons, and at still larger $Q$ (deep) inelastic interactions with quarks will become relevant. 
As shown in Refs.~\cite{Magill:2016hgc, Zhou:2019vxt}, the inelastic contribution to trident production is negligible (at most 1\%) in the energy range of our interest.

Kinematically, the flavor structure of neutrino tridents is limited by the mass-hierarchy of charged-leptons. 
Conservation of momentum and energy yields a threshold of initial neutrino energy for the production of a trident: 
\begin{align} E_{\nu}^\text{th}\approx  \frac{(m_\ell+m_\ell'+M_{\rm tgt})^2-M_{\rm tgt}}{2 M_{\rm tgt}}\, ,
\label{eq:thres}
\end{align}
where for incoherent scattering $M_{\rm tgt}=M_N$ is the nucleon mass $(\sim 1$ GeV$)$, and for coherent scattering $M_{\rm tgt}=M$ is the nuclear mass which may lead to an enhancement of the latter scattering regime. 
The threshold for various trident channels are listed in Table~\ref{tab:Thresholds} for three different nuclei, namely, argon, tungsten and iron, which are respectively the detector materials for DUNE, FASER$\nu$, and the T2K and MINOS near detectors. 
The larger tau mass increases the energy threshold and reduces the likelihood of tau-production via the trident, but it remains possible if the neutrino energy reaches above threshold. 
Moreover, incoherent scattering for tau tridents has a much higher threshold than coherent scattering, unlike the electron and muon tridents, where both processes have similar thresholds since $m_{e,\mu}/M_N,\,m_{e,\mu}/M\ll 1$.

\begin{table}[t!]
\centering
\def\arraystretch{1.5}
\begin{tabular}{|c|c|c|c|c|}
\hline 
  & \multicolumn{4}{c|}{  $E_\nu^\text{th}$ (GeV)}   \\ \cline{2-5}
\textsc{Trident process}  &  \textsc{Incoherent}&\multicolumn{3}{c|}{\textsc{Coherent} }\\ \cline{3-5}
& (proton)& $^{40}$Ar& $^{184}$W& $^{56}$Fe \\   
     \hline 
    $\nu_i N\to \nu_i N e^+ e^-$ & ~0.001~ &  ~0.001~ & ~0.001~&~0.001~\\
       $\nu_i N\to \nu_i N \mu^+ \mu^-$ &  0.24 &0.21& 0.21& 0.21\\
  $\nu_i N\to \nu_i N \tau^+ \tau^-$ &  10& 3.7  & 3.6& 3.7\\
$\nu_\mu N\to \nu_e N e^- \mu^+$ &0.11& 0.11 & 0.11&0.11\\
$\nu_\mu N\to \nu_\tau N \tau^- \mu^+$ &3.8& 1.9 &1.9& 1.9\\
$\nu_e N\to \nu_\tau N \tau^- e^+$ &3.5& 1.8 & 1.8& 1.8\\
\hline 
\end{tabular}
\caption{Threshold energies for various neutrino trident channels. For incoherent scattering, we utilize the proton mass as a proxy for the mean nucleon mass. }
\label{tab:Thresholds}
\end{table}

\begin{figure*}[t!]
    \centering  \includegraphics[width=0.8\textwidth]{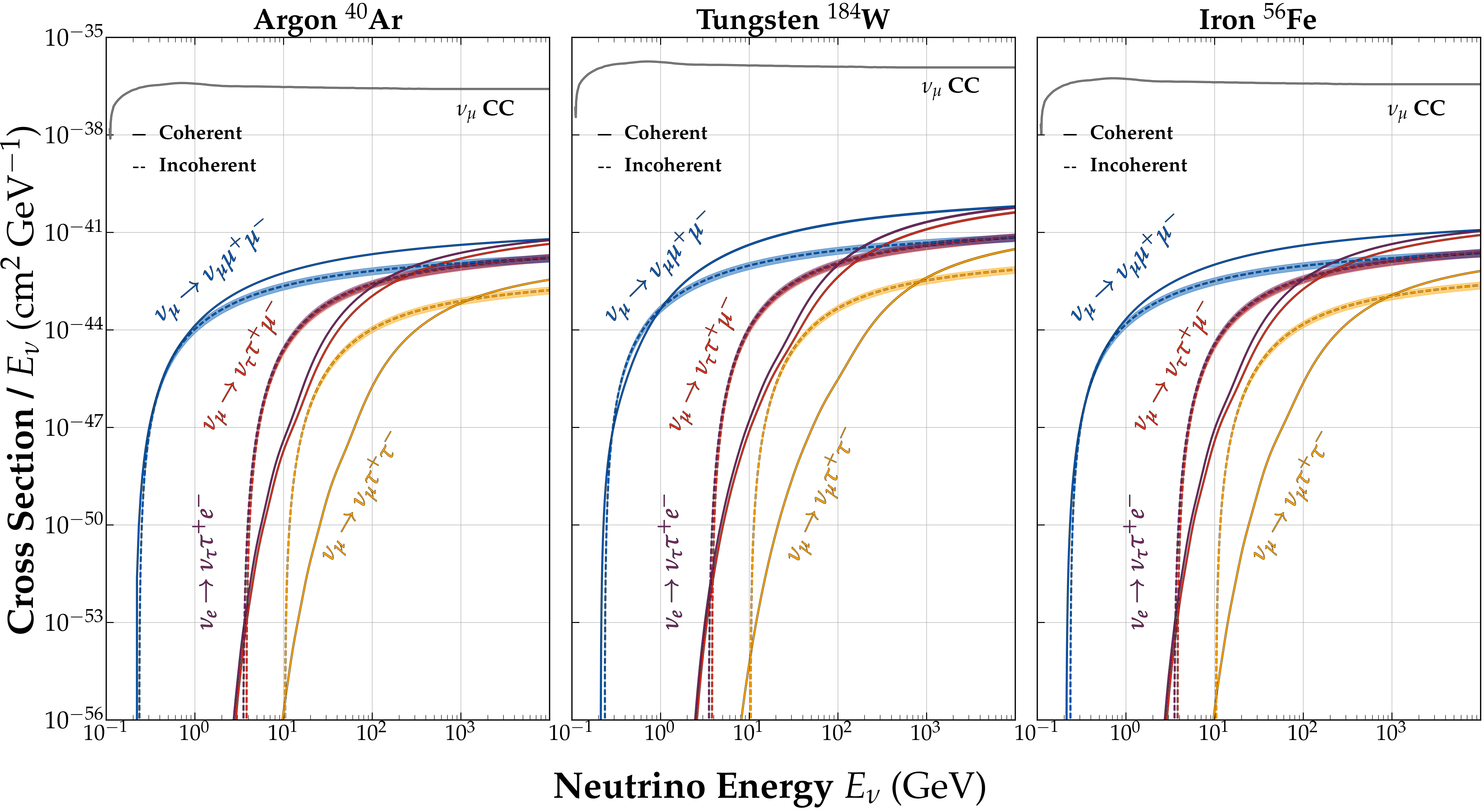}
    \caption{Coherent (solid) and incoherent (dashed) neutrino trident cross sections on different targets: argon (left), tungsten (middle) and iron (right). Note that $\nu_\mu  \to \nu_\tau \tau^+ \mu^-$ and $\nu_e \to \nu_\tau \tau^+ e^-$ have significant overlap.  
    The $\nu_\mu$ CC cross section is also shown for comparison. The antineutrino  cross sections are the same as for neutrinos.}
    \label{fig:xsec}
\end{figure*}
In both processes shown in Fig.~\ref{fig:NTP}, the neutrino interactions with $W$ or $Z$ give rise to two charged leptons, which then couple to the nucleus. 
Estimating the hadronic current has significant uncertainties, though it is independent of whether CC or NC interactions are generating the leptonic current in the trident process. 
Below $10^8$ GeV, the hadronic contributions arising from virtual weak boson-nucleus interactions and from mixing with the photon are negligible~\cite{Zhou:2019vxt}.

\subsection{Cross section calculation}
We calculate the full $2\to 4$ elastic scattering for both coherent and incoherent processes using the procedure described in Ref.~\cite{Ballett:2018uuc}, where
\begin{align}
    \label{eq:Xsec0}\frac{\mathrm{d} \sigma_{\nu X}}{\mathrm{d}Q^2 d\hat{s}}&= \frac{1}{32\pi^2} \frac{1}{\hat{s} Q^2} \times \\
    &\left[h_X^T(Q^2, \hat{s}) \sigma^T_{\nu\gamma}(Q^2, \hat{s})  +h_X^L(Q^2, \hat{s}) \sigma^L_{\nu\gamma}(Q^2, \hat{s})\right] \, ,\nonumber 
\end{align}
where $X=c,i$ denote the coherent and incoherent scattering regimes, respectively. $Q^2=-q^2$ is the photon virtuality, and $\hat{s}\equiv Q^2+(k_1+q)^2$, where $k_1$ is the four-momentum of the incoming neutrino. 
Eq.~\eqref{eq:Xsec0} factorizes the contributions to the trident process into transverse ($T$) and longitudinal ($L$) hadronic ($h_X^{L/T}$) and leptonic ($\sigma_{\nu \gamma}^{L/T}$) components. The functions $h_X^{L/T}$ are dimensionless flux factors including the nuclear (nucleon) form-factors $F(Q^2)/H(Q^2)$ for coherent (incoherent) scattering, as will be discussed below.

The leptonic components are calculated as cross sections between a neutrino and an off-shell photon,
\begin{align}
   \sigma_{\nu \gamma}^{L/T}= \frac{1}{2 (k_1+q)^2} \int \mathrm{d} \text{PS}_3 \frac{1}{2} \sum_\text{spins} |\mathcal{M}^{L/T}_{\nu \gamma}|^2  \, ,
\end{align}
where $\mathrm{d} \text{PS}_3$ is the Lorentz invariant three-body phase space, and $\frac{1}{2}\sum_\text{spins}|\mathcal{M}^{L/T}_{\nu \gamma}|^2$ is the spin-averaged matrix element for $\nu_\ell +\gamma_{L/T} \to \nu_{\ell'} +\ell' + \ell$, with a longitudinal or transversely polarized photon. We comment on the alternative Equivalent Photon Approximation (EPA) method for calculating $\sigma_{\nu \gamma}$ in Appendix~\ref{app:epa}.
\subsubsection{Coherent scattering}
In the coherent regime, the flux factors in Eq.~\eqref{eq:Xsec0} have the form
\begin{align}
    h_c^{T}(Q^2, \hat{s})&=8 Z^2 e^2
    \left( 1- \frac{\hat{s}}{2 E_\nu M} - \frac{\hat{s}^2}{4 E_\nu^2 Q^2}\right)|F(Q^2)|^2,
\end{align}
\begin{align}
    h_c^{L}(Q^2, \hat{s})&= 4Z^2 e^2\left( 1-\frac{\hat{s}}{4 E_\nu M}\right)^2 |F(Q^2)|^2,
\end{align}
where $E_\nu$ is the incoming neutrino energy, $M$ is the nuclear mass, and $F(Q^2)$ is the nuclear form factor for the specific target nucleus, details of which may be found in Appendix~\ref{app:ff}. 

\subsubsection{Incoherent scattering}
In the incoherent (diffractive) regime, the flux factors appearing in Eq.~\eqref{eq:Xsec0} have the form
\begin{align}
    h_i^{T}(Q^2, \hat{s})= 8e^2
   \bigg{[}\big{(} 1- &\frac{\hat{s}}{2 E_\nu M_N} - \frac{\hat{s}^2}{4 E_\nu^2 Q^2}\big{)}H_1^N(Q^2) \\
 & + \frac{\hat{s}^2}{8E_\nu^2M_N^2} H_2^N(Q^2)\bigg{]}, \\
     h_i^{L}(Q^2, \hat{s})= 4e^2
     \bigg{[} \big{(} 1- &\frac{\hat{s}}{4 E_\nu M_N} \big{)}^2H_1^N(Q^2)\\
   & - \frac{\hat{s}^2}{16 E_\nu^2 M_N^2}H_2^N(Q^2)\bigg{]}, 
\end{align}
where $M_N=m_{p,n}$ is the proton or neutron mass, and  $H_{1/2}^N(Q^2)$ are the nucleon form factors, defined as
\begin{align}
H_1^N(Q^2) &= |F_1^N(Q^2)|^2- \tau |F_2^N(Q^2)|^2,\label{eq:nucleon1} \\ H_2^N(Q^2)& = |F_1^N(Q^2)+F_2^N(Q^2)|^2,\label{eq:nucleon2} 
\end{align}
where $\tau = -Q^2/4M^2$. The functions $F_1^N(Q^2)$ and $F_2^N(Q^2)$ appearing in the nucleon flux factors via Eq.~\eqref{eq:nucleon1} and ~\eqref{eq:nucleon2} can be related to the electric and magnetic form-factors of the nucleons as
\begin{align}
    G_E^N(Q^2) &= F_1^N + \tau F_2^N(Q^2) 
    ,\\
    G_M^N(Q^2) &= F_1^N +  F_2^N(Q^2),
\end{align}
where the form-factors used in this work are outlined explicitly in  Appendix~\ref{app:ff}.

For this incoherent contribution to the cross section, Eq.~\eqref{eq:Xsec0} is further multiplied by a Pauli-blocking factor, derived by modeling the nucleus as an ideal Fermi Gas, assuming equal density of protons and neutrons,
\begin{align}
    \Theta(|\vec{q}|) = \begin{cases} 
      \frac{3}{2} \frac{|\vec{q}|}{2 k_F} - \frac{1}{2} \left(\frac{|\vec{q}|}{2 k_F} \right)^3 & {\rm for}~ |\vec{q}|<2 k_F \\
     1 & {\rm for}~ |\vec{q}|\geq 2 k_F 
   \end{cases} \, ,
\end{align}
where $\vec{q}$ is the spatial component of the momentum transfer to the nucleus, and $k_F$ is the Fermi momentum of the gas, taken to be $k_F=235$ MeV~\cite{Lovseth:1971vv}.

\subsection{Total cross sections for benchmark nuclei}

Our cross section results are shown in Fig.~\ref{fig:xsec} for a number of benchmark nuclei: where argon is relevant for DUNE, tungsten and iron for other accelerator neutrino facilities: FASER$\nu$, T2K INGRID near detector and MINOS/MINOS+ for contrast; see Appendix~\ref{app:exp}.

Following the methodology of Ref.~\cite{Altmannshofer:2019zhy}, we estimate the cross section uncertainties for coherent (incoherent) scattering as follows. We assume a $1\%$ ($3\%$) uncertainty for the nuclear (nucleon) form factors.\footnote{{This is the error budget for a normalization uncertainty. If we consider an uncertainty in the shape of the form factor, we do not expect significant impact  on our final results as the missing-energy nature of the neutrino final state will smear any potential effects.}} Higher order QED corrections scale as $\sim Z \alpha_\mathrm{EM} / 4\pi$; we assume a conservative $3\%$ ($3\%$) uncertainty for argon, $6\%$ ($3\%$) uncertainty for tungsten and $4\%$ ($3\%$) uncertainty for iron. Our choice of weak mixing angle, $\sin^2\theta_W = 0.23129$, also leads to a $5\%$ ($5\%$) uncertainty. Additionally, given our simplified nuclear model of a Fermi gas, we include a conservative $30\%$ uncertainty for incoherent scattering. We add all our uncertainties in quadrature to get a total estimated uncertainty of $6\%$ ($31\%$) for argon, $8\%$ ($31\%$) for tungsten, and $6\%$ ($31\%$) for iron. 

Due to lower neutrino energy threshold (see Table~\ref{tab:Thresholds}), the coherent contribution dominates the trident cross section at low neutrino energies.
This is clearly evident for the tau trident processes for which there is a larger separation in the energy thresholds between the coherent and incoherent processes.  
As the neutrino energy increases, the minimum photon momentum-transfer, $Q^2$, necessary to generate the final state will decrease. 
As $Q^2$ increases, the nuclear form factor decreases and the hadronic form factors come to dominate as the neutrino can resolve more of the nuclear structure (see {Appendix~\ref{app:ff}). 
This leads to an initial crossing point where the cross section is dominated first by the coherent scattering and then by incoherent.  For larger incoming neutrino energies, the contribution from coherent scattering is proportional to $Z^2$ (where $Z$ is the atomic number), and the contribution from incoherent scattering scales with the number nucleons, i.e. proportional to $\sim 2 Z$. As the neutrino energy increases further, the enhancement from the $Z^2$ dependence of the coherent scattering will again dominate over the $2Z$ evolution of the incoherent piece, leading to a second crossing. This is most prominent for the tau trident processes in Fig.~\ref{fig:xsec}.

These two crossings are more pronounced in the tau trident case because they occur at much larger and more distinct neutrino energies. This is because of the presence of the tau, whose mass influences the momentum transfer $Q^2$. This pushes the crossing points to occur at larger neutrino energies and, therefore, there is a considerable region of energies sampled by the DUNE flux where incoherent production dominates the tau trident. We see from Fig.~\ref{fig:xsec} that coherent scattering dominates the muon trident much more rapidly as evolved with $E_\nu$.

\begin{figure*}[t!]
    \centering
    \includegraphics[scale=0.141]{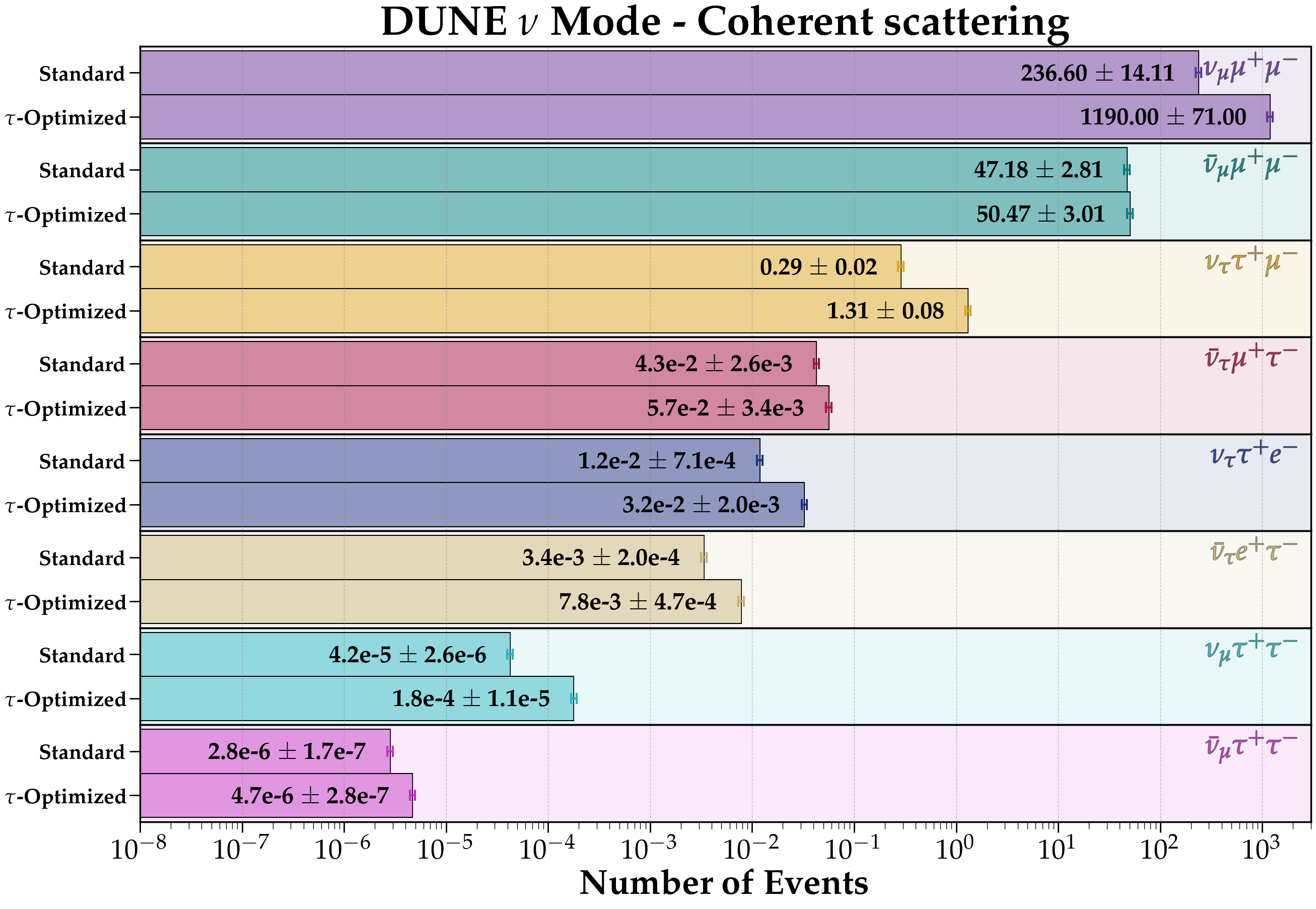}
\includegraphics[scale=0.141]{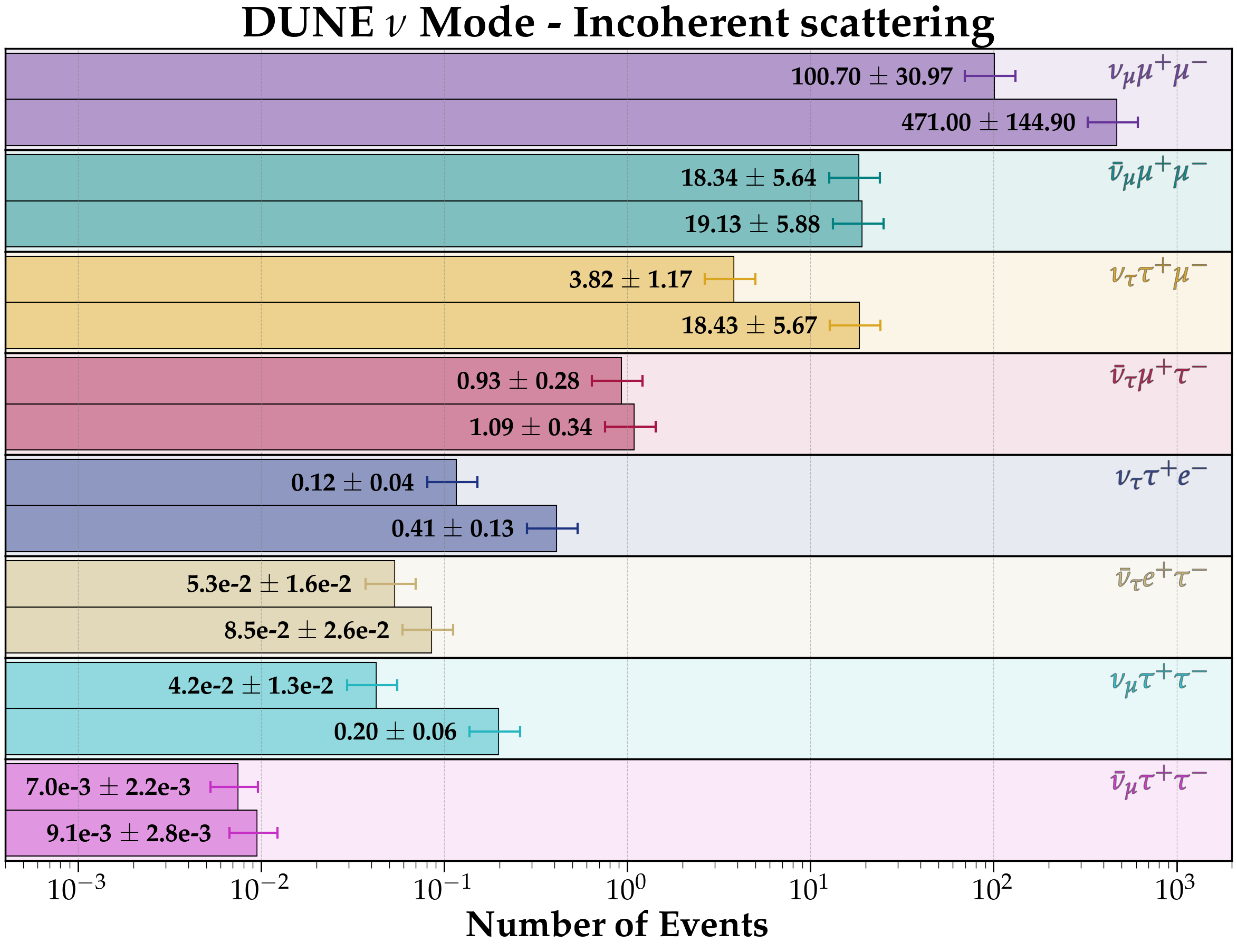} 
    \caption{Number of trident events from coherent (left) and incoherent (right) scattering at DUNE near detector with 67 ton of argon and $3.3\times10^{21}$ POT in the forward horn current mode for both standard and tau-optimized fluxes.}
    \label{fig:DUNE}
\end{figure*}

\section{Neutrino tridents at DUNE} \label{sec:III}
The DUNE experiment~\cite{DUNE:2015lol} involves the production of an intense muon-neutrino beam from impinging 120 GeV proton beams on a graphite target, and placing detectors both near and far from the beam origin. 
These $\nu_\mu$'s with energies ranging up to several GeVs will then travel along the beamline and either interact with the detector or oscillate to other neutrino flavors.
The posited length $L = 574$ m between the beam origin and the ND is not sufficient for $\nu_\tau$ production via standard neutrino oscillations for GeV-energy neutrinos. 
Although heavy charmed meson ($D$, $D_s$) decays could directly give $\nu_\tau$'s, their production rate is expected to be very small for a 120 GeV proton beam. 
While the observation of a signal consistent with the presence of $\nu_\tau$ is often considered `anomalous' and a prime target for BSM searches~\cite{Coloma:2017ptb, Bakhti:2018avv, DeGouvea:2019kea, Ghoshal:2019pab, Machado:2020yxl, Coloma:2021uhq,Bigaran:2022giz, Dev:2023rqb}, neutrino tridents can be a significant source of tau leptons in the near detector, as we will see now.

With the trident cross sections calculated above, we obtain the expected number of events at an accelerator neutrino experiment following the methodology of Ref.~\cite{Altmannshofer:2019zhy}:  
\begin{align}
    N_{\text{trident}} = \Phi \sigma_\text{conv} \frac{M_\text{det}}{M} N_\text{POT} \, , 
    \label{eq:event}
\end{align}
where $N_\text{POT}$ is the number of protons on target, $\Phi$ is the relevant integrated neutrino (or antineutrino) flux, $M_{\rm det}$ is the detector mass, $M$ is the nuclear mass, and $\sigma_\text{conv}$ is the cross section convoluted with the normalized neutrino flux: 
\begin{align}
    \sigma_\text{conv} = \int \frac{1}{\Phi} \frac{\dd{}\Phi}{\dd{}E_\nu} \sigma_{\nu X} \dd{}E_\nu\, . 
\end{align}
Here $\sigma_{\nu X}$ is the trident cross section. The relevant neutrino fluxes at various detectors considered here can be found in Appendix~\ref{app:flux}. 
For DUNE, we consider the standard neutrino flux, as well as the tau-optimized configuration with more energetic neutrinos~\cite{DUNE:2020lwj, Fields}. 

In Fig.~\ref{fig:DUNE}, we show the number of trident events, for all $\nu_\mu$-initiated tau tridents as well as $\nu_\mu \to \nu_\mu \mu^+ \mu^-$ and $\nu_e \to \nu_\tau \tau^+ e^-$, at DUNE-ND with 67 ton of argon and $3.3\times 10^{21}$ POT in the forward horn current, or neutrino mode (see Appendix~\ref{app:exp} 
for antineutrino mode). 
The uncertainties shown come from the cross section uncertainties outlined above. 
For simplicity, we do not take into account uncertainties in the neutrino fluxes, which can in principle be reduced with other measurements (e.g., neutrino-electron scattering) at DUNE-ND.
Since experimentally one may be able to distinguish between coherent and incoherent scattering by observing the hadronic activity, we show their corresponding number of events separately.
For the ease of cross check with previous literature~\cite{Ballett:2018uuc, Altmannshofer:2019zhy}, we report the dimuon trident events as well, although our main focus is on the taus. For instance, our event numbers for the dimuon tridents roughly match with those given in Ref.~\cite{Altmannshofer:2019zhy}, after taking into account the differences in the flux and detector mass used there from the older DUNE configuration.
As for the tau tridents, in total, we expect $4.11 \pm 1.17$ ($0.97 \pm 0.28$) events in the standard mode and $19.74 \pm 5.67$ ($1.15 \pm 0.34$) events in the tau-optimized mode for the $\nu_\mu \to \nu_\tau \tau^+ \mu^-$ ($\bar{\nu}_\mu \to \bar{\nu}_\tau \tau^- \mu^+$) trident running in the forward-horn configuration. Most of these events come from incoherent scattering due to DUNE's energy spectrum. 
This is in sharp contrast with the dimuon trident, where the coherent contribution is the dominant one. 
The tau-optimized mode gives more events because the flux peaks at higher energies, making it easier to pass the energy threshold for the tau tridents. 
For ditau tridents, we expect less than one event even in the tau-optimized mode. For completeness, we also report the numbers for $\nu_e \to \nu_\tau \tau^+ e^-$ and $\nu_\mu \to \nu_\mu \tau^+ \tau^-$ as well as their antineutrino counterparts in Fig.~\ref{fig:DUNE}. In Appendix~\ref{app:exp} 
we also show the number of trident events for FASER$\nu$, T2K INGRID near detector and MINOS/MINOS+ for comparison.

\section{\texorpdfstring{Tau Identification from Tridents at DUNE}{nutau  Event Identification}} \label{sec:ID}

Detecting tau leptons in a realistic neutrino experimental setup like DUNE is notoriously challenging. 
First of all, LArTPCs cannot identify taus on an event-by-event basis due to the very short displaced vertex.
Besides, the tau production cross section, either from intrinsic beam $\nu_\tau$ or tridents, is very low, making virtually any experimental setup statistics-limited for tau observation. 
Tau reconstruction suffers from large backgrounds induced by the copious beam $\nu_\mu$ component.
Misreconstruction of final state particles could significantly degrade the signal-to-background ratio.
This has been considered a major challenge in $\nu_\tau$ detection at DUNE in the context of BSM studies~\cite{DeGouvea:2019kea,Ghoshal:2019pab,Giarnetti:2020bmf,Coloma:2021uhq}. 
However, recent developments based on advanced machine-learning algorithms hold some promise~\cite{TautagAdam, nutautalk, neutrinotalk, SnowmassTalk}.

The basic idea is to separate the tau signal events from backgrounds using kinematic differences, similar to those used in NOMAD~\cite{NOMAD:2001xxt}. 
Reference~\cite{SnowmassTalk} ranked six kinematic variables providing the highest signal-to-background for tau reconstruction. 
The ratio of the transverse momenta, $R^T_{\rm Miss}=p^T_{\rm Miss}/(p^T_{\rm Miss}+p^T_\mu)$,  was identified as the rank-1 discriminator. 
This ratio $R^T_{\rm Miss}$ is particularly relevant for our tau trident signal which always comes with additional missing energy from neutrinos, and therefore, we expect a larger $R^T_{\rm Miss}$ for the signal compared to the background.

\begin{figure}[t!]
    \centering
    \includegraphics[width=0.49\textwidth]{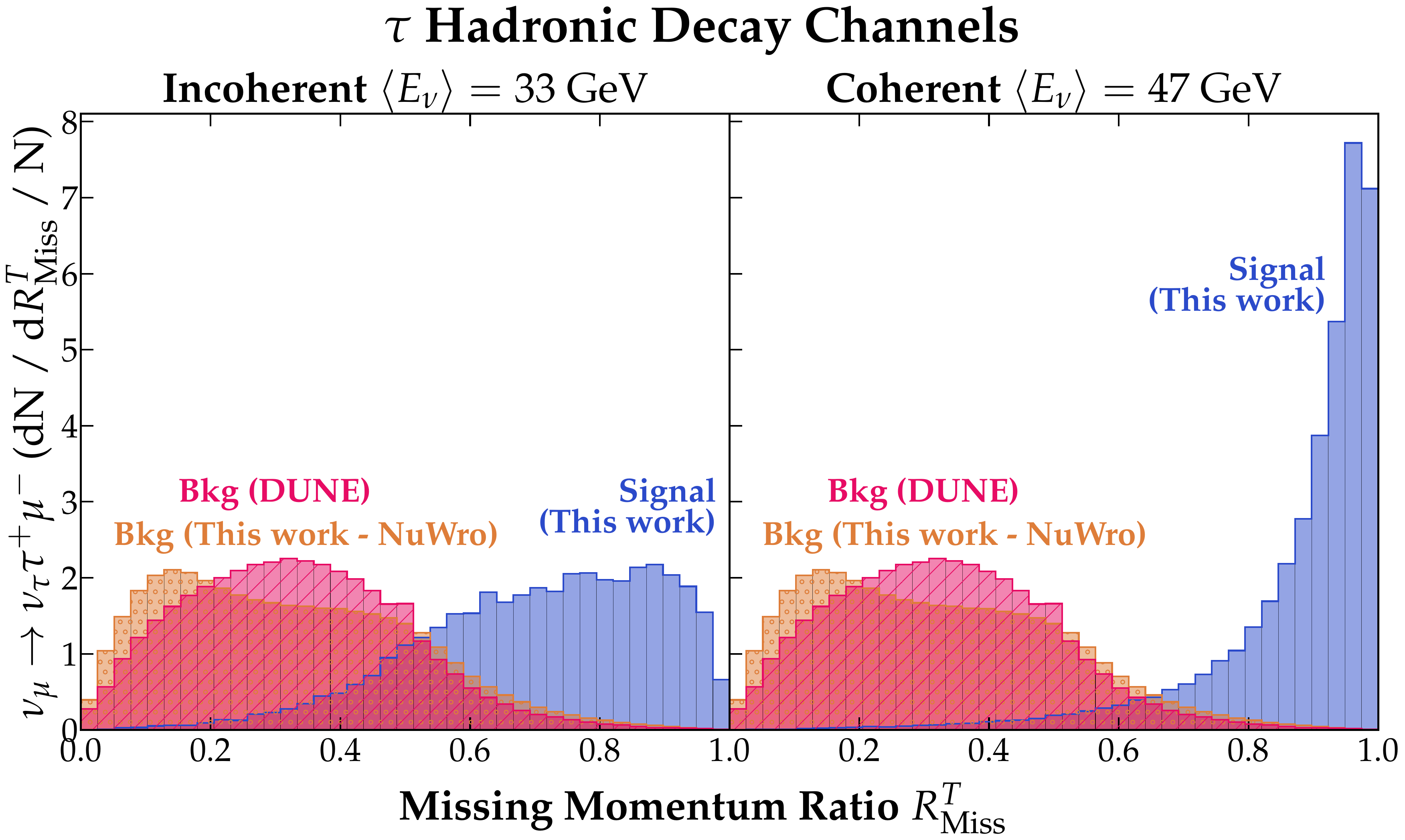}
    \caption{Missing transverse momentum ratio distributions for the $\nu_\mu \to \mu^-\nu_\tau \tau^+ \to \mu^- \nu_\tau \bar{\nu}_\tau + {\rm hadrons} $. The incoherent (left) and coherent (right) contributions are shown separately for two benchmark neutrino energies corresponding to the peaks in DUNE event spectra.
    The dominant $\nu_\mu$ CC background is obtained from the {\tt NuWro} event generator (orange) and the official {\tt GENIE/GEANT4}-based DUNE simulation (pink). The latter is extracted from Ref.~\cite{SnowmassTalk} (see also Refs.~\cite{nutautalk, neutrinotalk}).
    } 
    \label{fig:RMiss_1tau}
\end{figure} 

In Fig.~\ref{fig:RMiss_1tau}, we show the $R^T_{\rm Miss}$ distributions for the $\mu\tau$ trident with incoherent (left) and coherent (right) contributions in blue for two benchmark average neutrino energies derived from their event distributions (Fig.~\ref{fig:event_dist} in Appendix~\ref{app:dist}). 
Here we consider the hadronic decays of tau, due to their larger branching ratio (65\%), compared to the leptonic decays (35\%). We have simulated the hadronic tau decays using {\tt PYTHIA 8.3}~\cite{Bierlich:2022pfr} and subsequently boosted them to our trident frame.
For comparison, we show the dominant background from $\nu_\mu$ CC
extracted from the works of Refs.~\cite{TautagAdam, nutautalk, neutrinotalk,SnowmassTalk}, which used the official {\tt GENIE/GEANT4}-based DUNE simulation~\cite{Andreopoulos:2009rq,GEANT4:2002zbu,GEANT_physicsmanual}.
We also have generated the same background using the {\tt NuWro} event generator~\cite{Golan:2012rfa}.
{To account for detector effects, we smeared the 3-momenta and polar angles of particles as follows. 
For coherent scattering off argon, 
we smear $p^T_\text{Miss}$ by $10\%$ on each component. 
For incoherent scattering, $p^T_\text{Miss}$ is inferred instead from the visible outgoing particles (we treat neutrons as invisible).
We assume a 2\% (10\%) energy and $2^\circ$  ($10^\circ$) angular smearing for muons (other particles).
}

Although the DUNE and {\tt NuWro} distributions differ by $\sim$20\%, they both shrink at larger $R^T_{\rm Miss}$ where most of the signal is. 
{To estimate the impact of  $R^T_\text{Miss}$ as a discriminator, we generate $R^T_\text{Miss}$ distributions for neutrino energies ranging from $5-\qty{125}{\giga\electronvolt}$. 
We then applied an $R^T_\text{Miss} \geq 0.6$ cut and extracted the percentage of signal events that pass this cut at a given $E_\nu$, producing an energy-dependent efficiency. 
Including this efficiency, the expected number of  coherent scattering $\mu\tau$ trident events at DUNE ND in neutrino mode changed from $0.29\pm0.02$ to $0.28\pm0.02$ (standard beam) and from $1.31\pm0.08$ to $1.24\pm0.07$ (tau-optimized beam). 
For incoherent scattering, the corresponding numbers changed from $3.82\pm1.17$ to $2.88\pm0.88$ and from $18.43\pm5.67$ to $13.76\pm4.23$.
}
While not a full analysis, this serves to illustrate the point that discriminators like $R^T_{\rm Miss}$ could mitigate the large $\nu_\mu$ CC background, enhance the signal-to-background ratio, and thus potentially allow us to observe the handful of tau trident events at DUNE-ND. 
A detailed experimental feasibility study is left as future work. Moreover, high-energy and high-intensity neutrino beams available at future colliders could provide another promising avenue for tau tridents; see e.g.~Ref.~\cite{Bojorquez-Lopez:2024bsr}.


\section{Conclusions} \label{sec:con}
Any potential appearance of tau leptons at near detectors in accelerator-neutrino experiments has been associated with BSM physics in the literature. Therefore, it is of paramount importance to delineate all potential SM backgrounds for tau events. In this work, we identified one such SM source of tau lepton production at DUNE-ND, namely, tau tridents. 
To this effect, we performed for the first time a detailed study of tau tridents at DUNE-ND, and contrary to the common lore, found that they could yield a non-negligible event rate, especially with the tau-optimized flux configuration. 
We calculated the tau trident cross section using the full $2\to 4$ scattering in order to obtain an accurate prediction of the event rate and found an intricate interplay between coherent and incoherent scattering which is distinct for tau tridents from electron and muon tridents. 
Lastly, we have identified a promising venue to enhance the poor signal-to-background ratio using kinematic variables, potentially leading to the observability of tau tridents for the first time in the DUNE experiment.
We hope that these findings will motivate a dedicated experimental feasibility study of tau tridents at accelerator neutrino facilities. 


\acknowledgments
We would like to thank Wolfgang Altmannshofer, Alex Sousa and Bei Zhou for useful discussions. We also thank Toni M{\"a}kel{\"a} and Sebastian Trojanowski for useful correspondence on the FASER$\nu$ trident analysis. This manuscript has been authored in part by Fermi Research Alliance, LLC under Contract No. DE-AC02-07CH11359 with the U.S. Department of Energy, Office of Science, Office of High Energy Physics. This work of BD was partly supported by the U.S. Department of Energy under grant No. DE-SC 0017987. DLG is supported by an MCSS Graduate Fellowship. 

\appendix 

\section{Equivalent Photon Approximation Comparison}
\label{app:epa}
Here we remark on an alternative approach for calculating the $\sigma_{\nu \gamma}$ component of the cross section in Eq.~\eqref{eq:Xsec0}, 
utilizing the equivalent photon approximation~(EPA)~\cite{Belusevic:1987cw}. 
To date, the only tau trident estimates for DUNE are given in Ref.~\cite{Magill:2016hgc}, which employed the EPA, and deemed these processes inaccessible at DUNE energies.\footnote{Note also that the projected neutrino fluxes used in Ref.~\cite{Magill:2016hgc} have since been updated to those we employ in this work, and so the exact projected event numbers should not be directly compared.} 
In the EPA, the infrared pole of the propagator for on-shell photon exchange between the lepton and nucleus is assumed to yield a dominant contribution to the cross section. 
As $Q^2\to 0$, the transverse cross section approaches that of an on-shell photon, and the longitudinal component vanishes. 
Thus, when applying the EPA, the longitudinal contributions in Eq.~\eqref{eq:Xsec0} 
are omitted, and the total cross section scales roughly as $\sigma\sim G_F^2 E_\nu Q_\text{max} \log [2E_\nu Q_\text{max}/(m_\ell+ m_\ell')^2]$~\cite{Belusevic:1987cw, Magill:2016hgc}, where $G_F$ is the Fermi constant. 

The EPA method gives a good estimate for some neutrino trident cross sections when coherent scattering dominates~\cite{osti_4832757}. It overestimates the cross sections of the dielectron trident, but is a good approximation for the coherent contribution to dimuon tridents~\cite{Ballett:2018uuc, Zhou:2019vxt}. 
Ref.~\cite{Ballett:2018uuc} establishes that for accelerator neutrino energies, EPA is not valid for tridents in regions where the virtual photon contributions are considerable, and where the incoherent scattering contributions are non-negligible -- as we see is true for tau-containing tridents from our results in the main text.
\section{Form Factors}
\label{app:ff}
Here we outline the utilized nuclear and nucleon form factors, used for the calculation of coherent and diffractive neutrino trident scattering, respectively. For nuclear form factors, we highlight a series of benchmark target nuclei relevant not only for DUNE but also for the other experiments for which we provide event estimates in Appendix~\ref{app:exp}.
\subsection{Nuclear form factors }
\noindent For argon and iron, we have used
\begin{align}
    F(Q^2) = \int \mathrm{d}r \: r^2\frac{\sin(Qr)}{Qr} \rho(r)  \, ,
\end{align}
where $\rho$ is the spherically symmetric charge distribution of the nucleus, normalized as 
\begin{align}
\int \mathrm{d}r \: r^2\rho(r)=1 \, ,
\label{eq:norm}
\end{align}
such that $F(0)=1$. 
We rely on the nuclear charge density distributions fitted to elastic electron scattering data~\cite{DeVries:1987atn}. 
More specifically, we use the three-parameter Fermi charge distributions from Ref.~\cite{DeVries:1987atn}. 

For tungsten, we use the Woods-Saxon form factor~\cite{Woods:1954zz, Fricke:1995zz, Jentschura:2009mb}
\begin{align}
    F(Q^2) = \frac{1}{\int \mathrm{d}^3 r \rho(r)} \int \mathrm{d}^3 r \rho(r) \exp(-i \vec{q}\cdot \vec{r}) \, ,
\end{align}
where the charge distribution follows the functional form
\begin{align}
    \rho(r)=\frac{{\cal N}}{1+\exp{\left(\frac{r-r_0}{\sigma_0}\right)}} \, ,
\end{align}
with $r_0=1.126~{\rm fm}\:A^{1/3}$ and $\sigma_0=0.523$ fm. Here ${\cal N}$ is a normalization factor that is determined from the condition in Eq.~\eqref{eq:norm}.
This form factor can be expressed analytically using a symmetrized Fermi function as outlined in Ref.~\cite{Ballett:2018uuc},
\begin{align}
    F(Q^2) = \frac{3\pi\sigma_0}{r_0^2+\pi^2\sigma_0^2} \frac{\pi\sigma_0\coth(\pi Q\sigma_0)\sin(Qr_0) - r_0\cos(Qr_0)}{Qr_0\sinh(\pi Q\sigma_0)}.
\end{align}
This is the final functional form used in the calculation of the tungsten form factor. 

In Fig.~\ref{fig:form_factor_argon}, we show the nuclear form factors for argon, tungsten and iron. It is clear that all of them drop sharply as the momentum transfer increases, which is expected for coherent scattering.

\begin{figure*}[t!]
    \centering
    \includegraphics[width=0.32\textwidth]{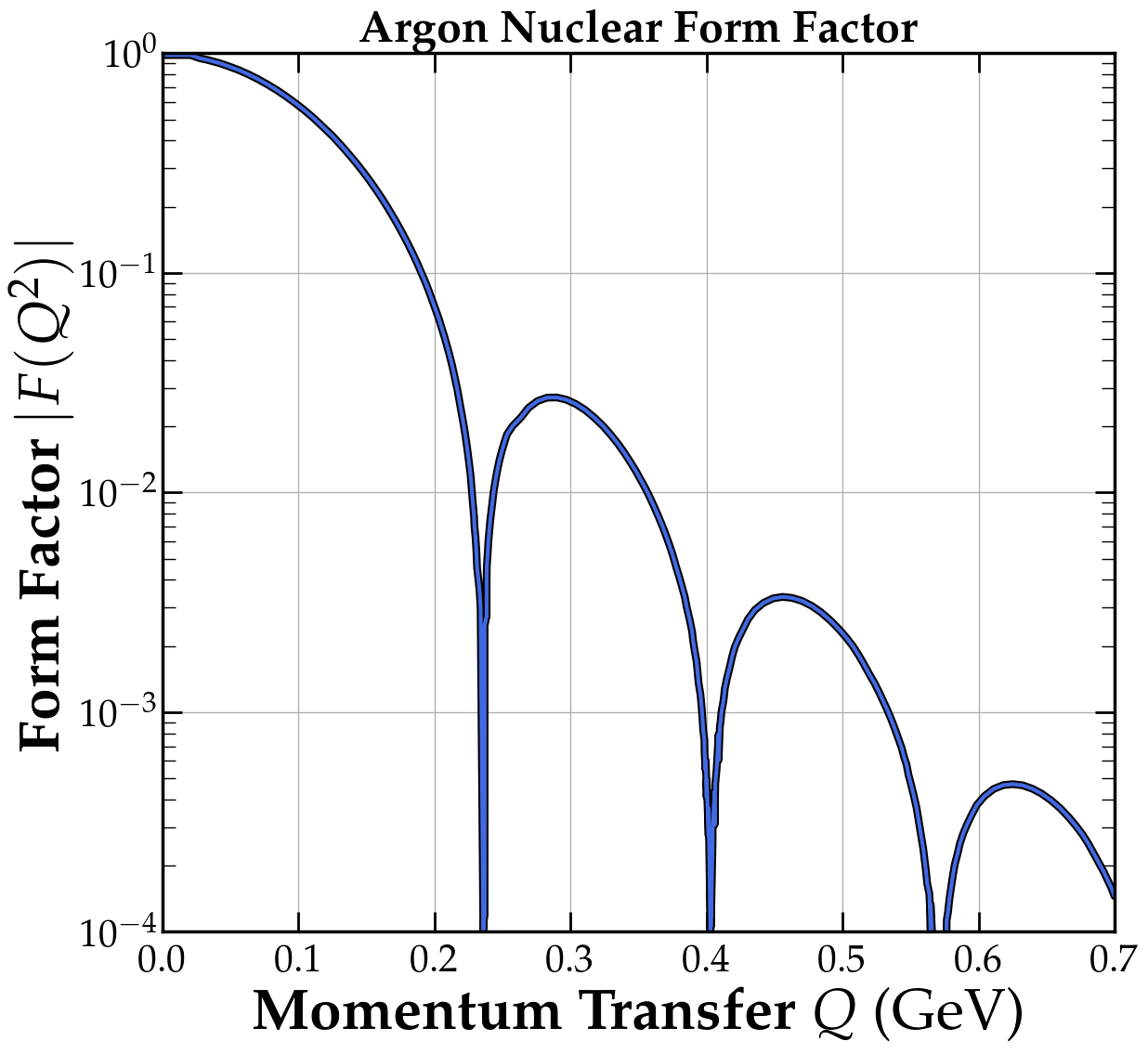}
     \includegraphics[width=0.32\textwidth]{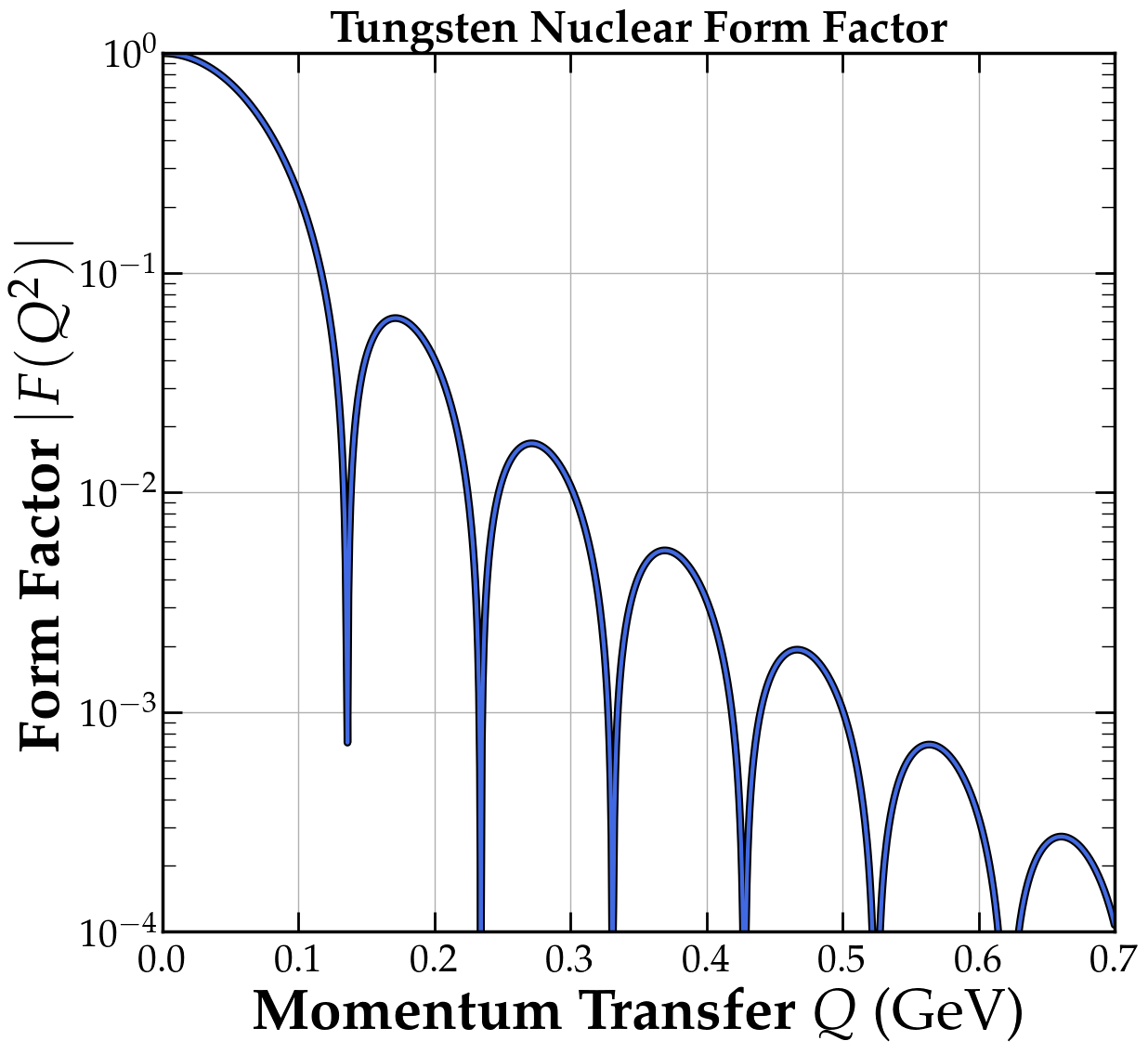}
     \includegraphics[width=0.32\textwidth]{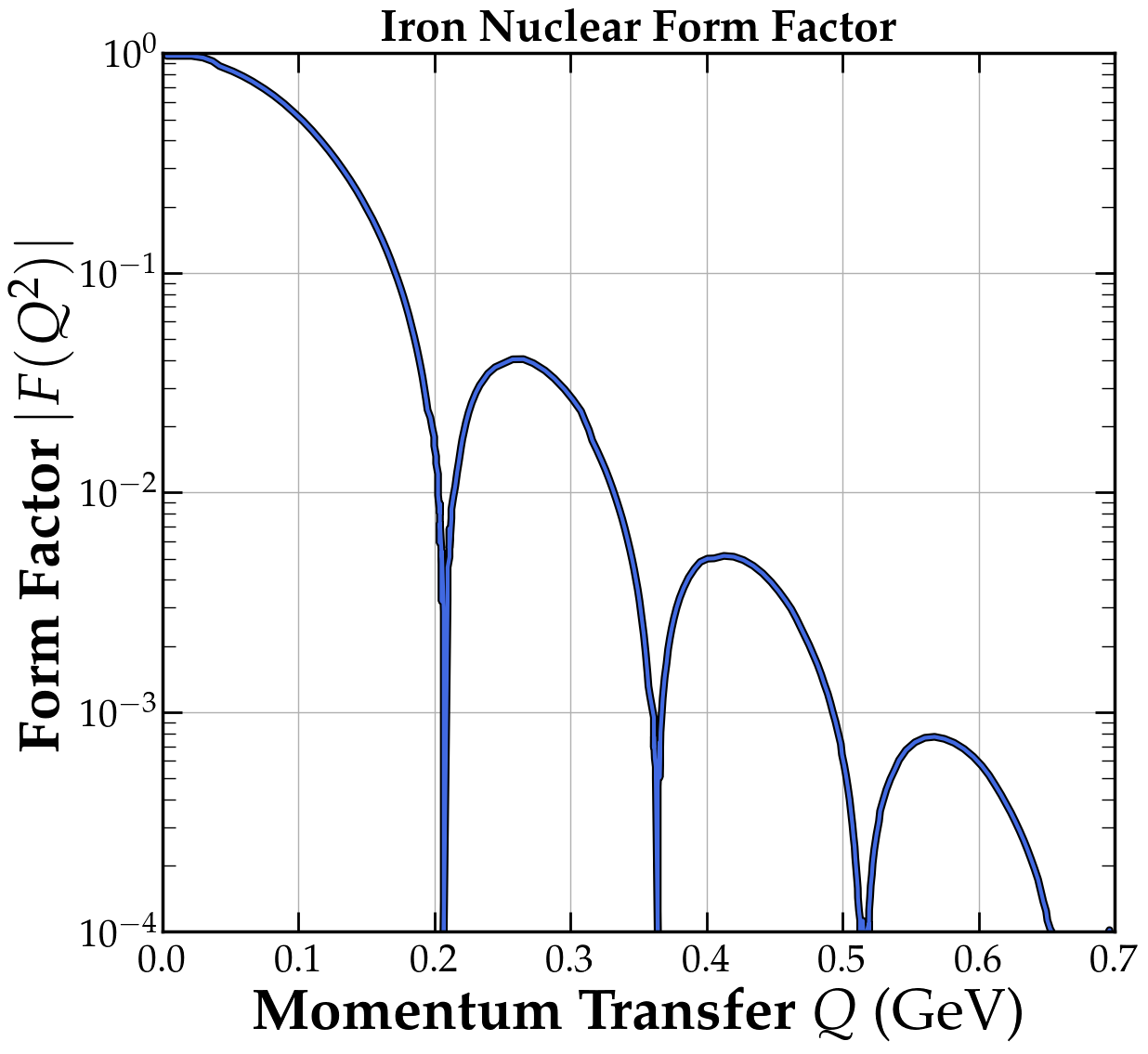}
    \caption{Nuclear form factors for argon (left), tungsten (middle) and iron (right).}
    \label{fig:form_factor_argon}
\end{figure*} 

\begin{figure}[t!]
    \centering
\includegraphics[width=0.4\textwidth]{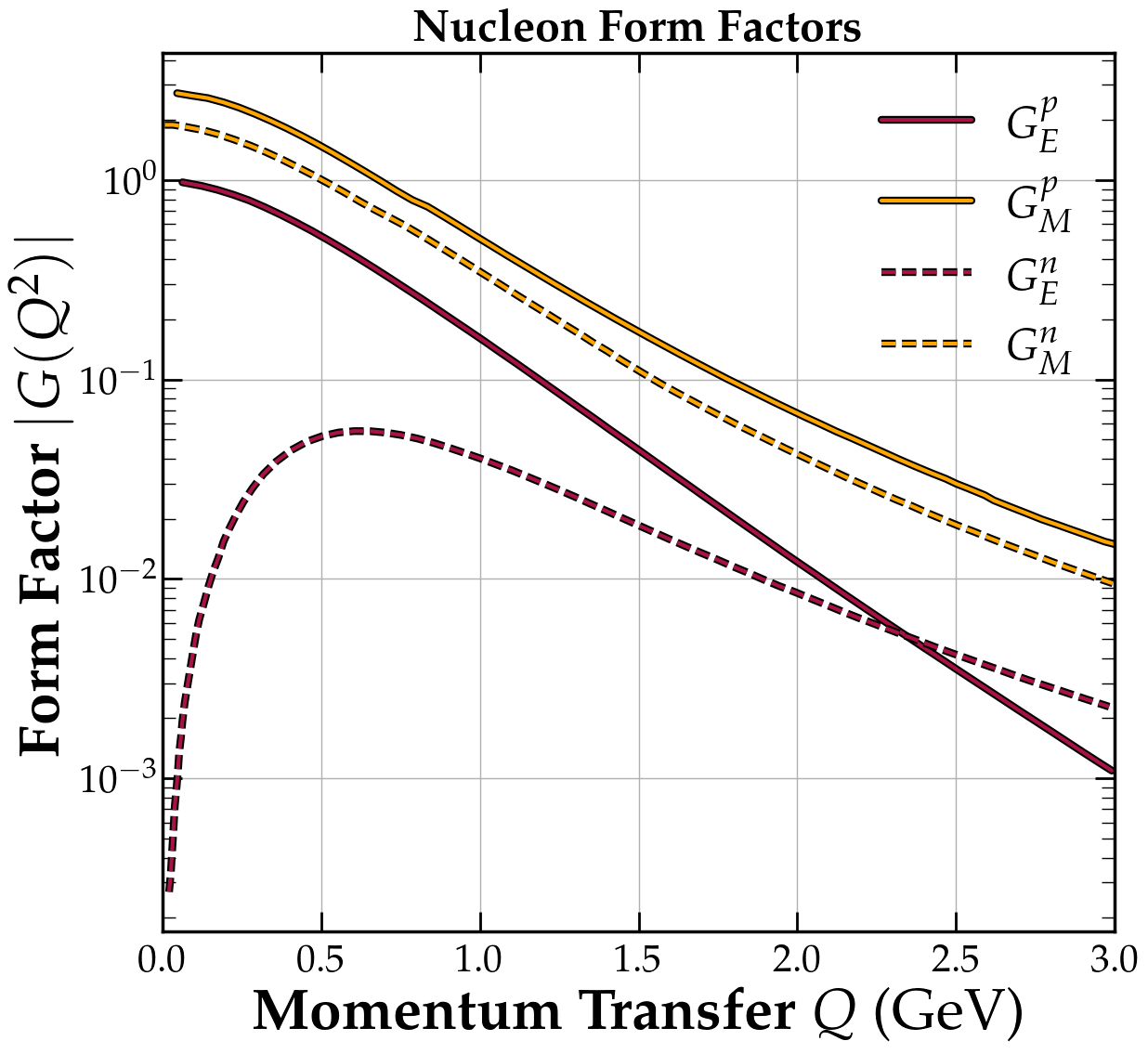}
    \caption{Electric and magnetic form factors of the proton and neutron. } \label{fig:form_factor_nucleons}
\end{figure} 

\subsection{Nucleon form factors}

While Ref.~\cite{Ballett:2018uuc} utilizes a simple dipole parameterization for $G_{E,M}^N(Q^2)$, we utilize those from Ref.~\cite{Alberico:2008sz} (see also Ref.~\cite{Ye:2017gyb} for a recent reevaluation). The proton form factors were obtained from fits to the electron-proton elastic scattering cross section and polarization transfer measurements. The neutron form factors were obtained from fits to electron-nucleus (mostly $^2$H and $^3$He) scattering data. Our results for the electric and magnetic form factors are shown in Fig.~\ref{fig:form_factor_nucleons}. Note that the electric form factor of the neutron goes to zero only at zero momentum transfer, whereas in the simple dipole approximation, it vanishes for all $Q$ values.


\section{Neutrino Fluxes}
\label{app:flux}
\begin{figure*}[t!]
\centering
\includegraphics[width=0.48\textwidth]{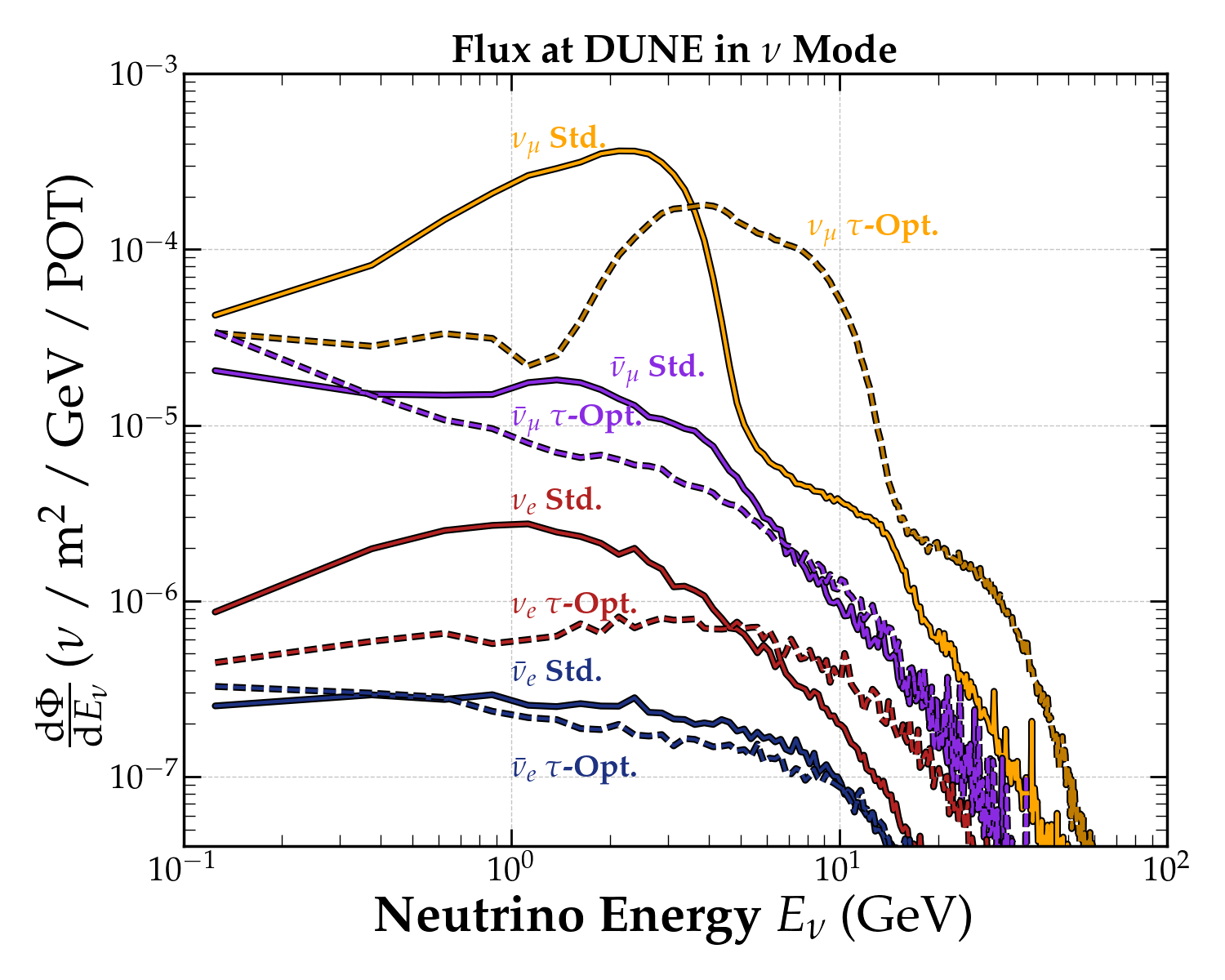}
\includegraphics[width=0.48\textwidth]{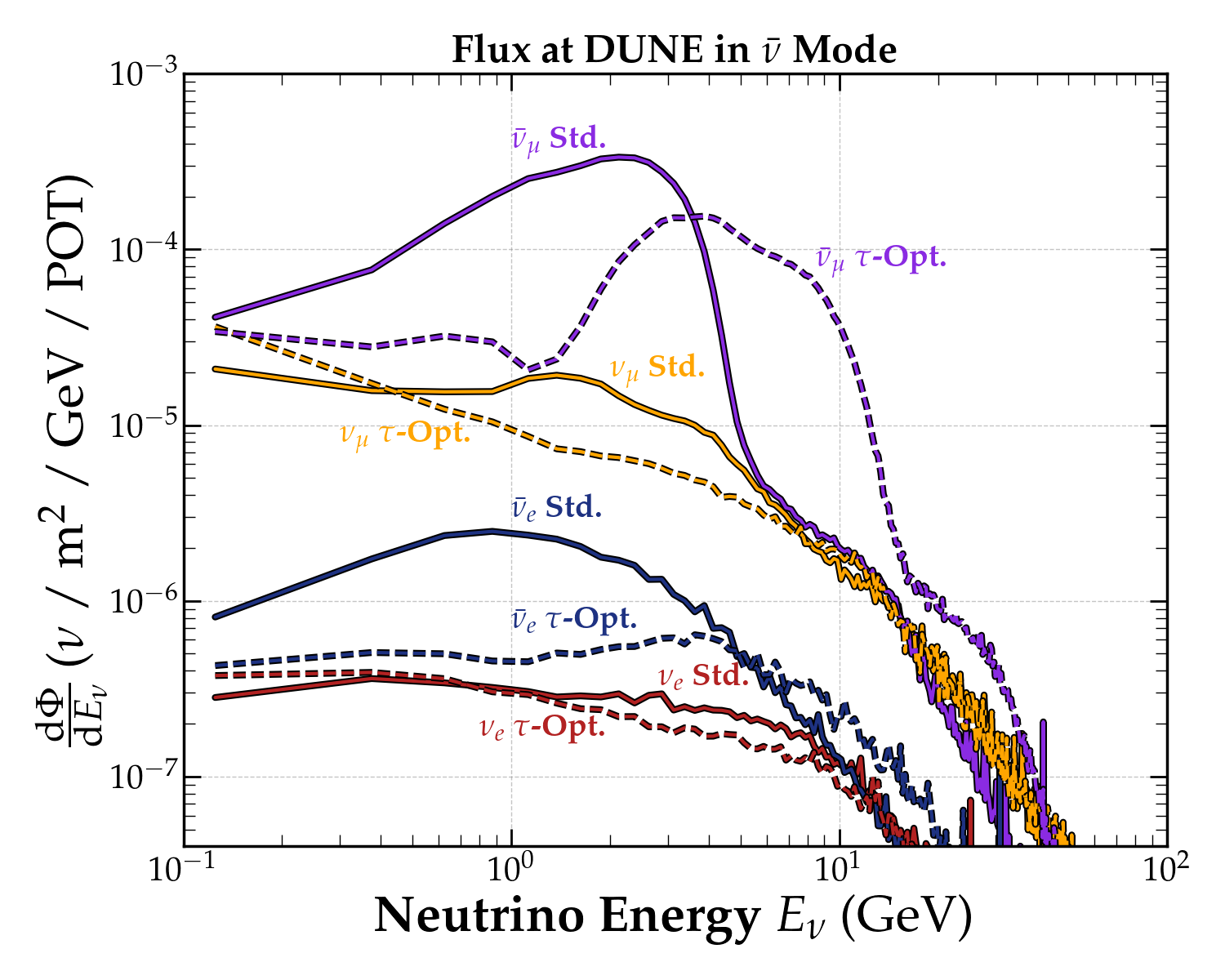}
\includegraphics[width=0.48\textwidth]{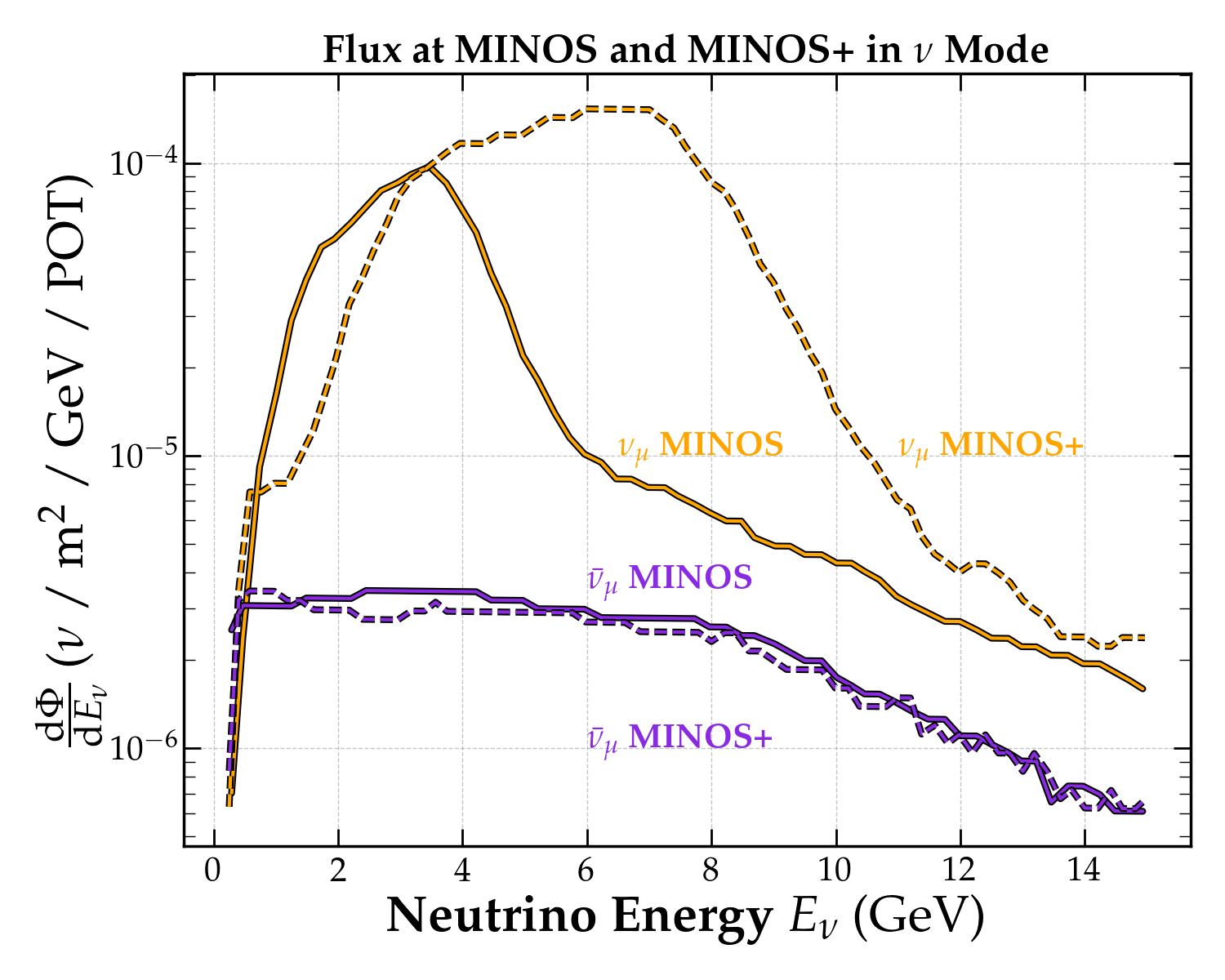}
\includegraphics[width=0.48\textwidth]{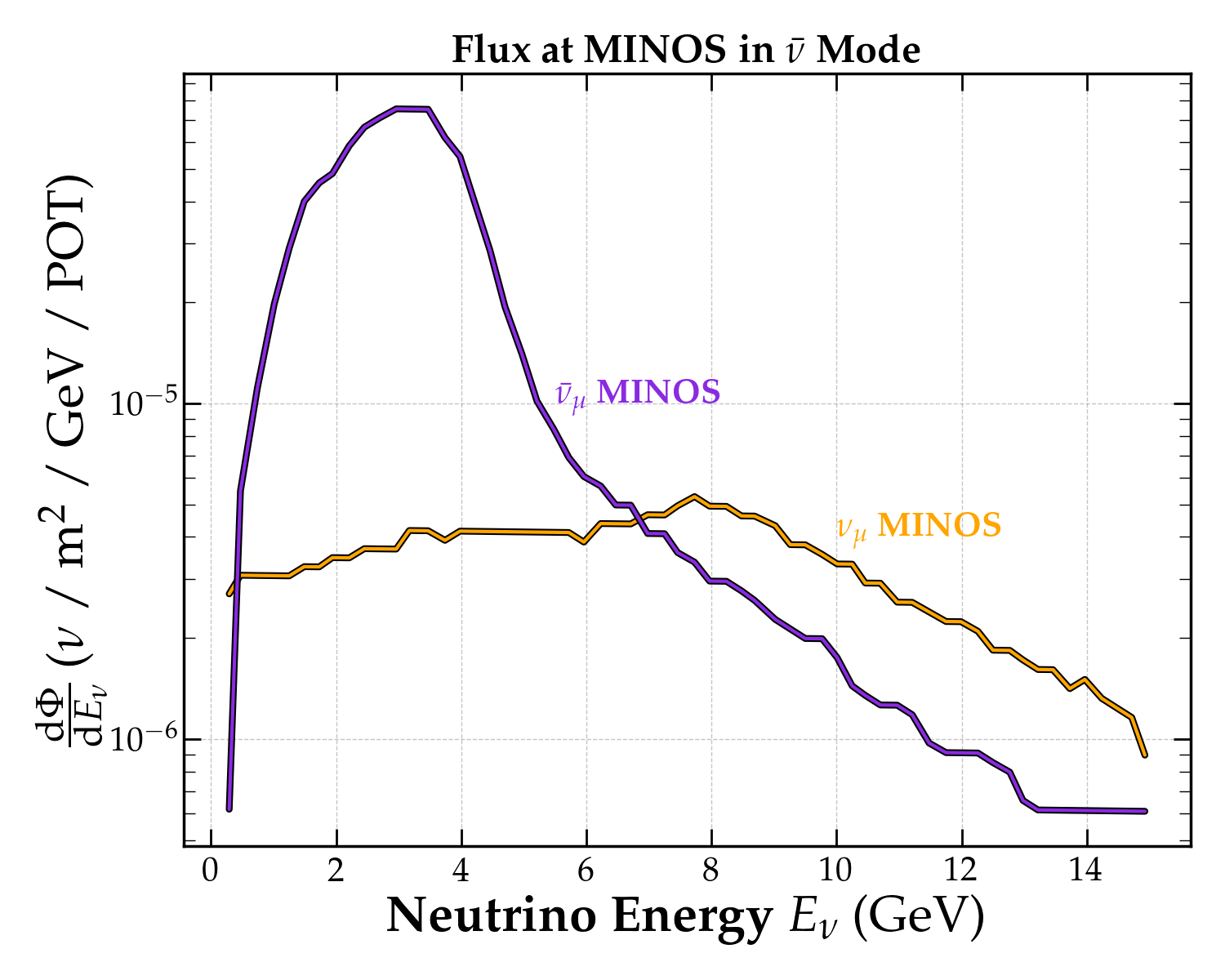}
\includegraphics[width=0.48\textwidth]{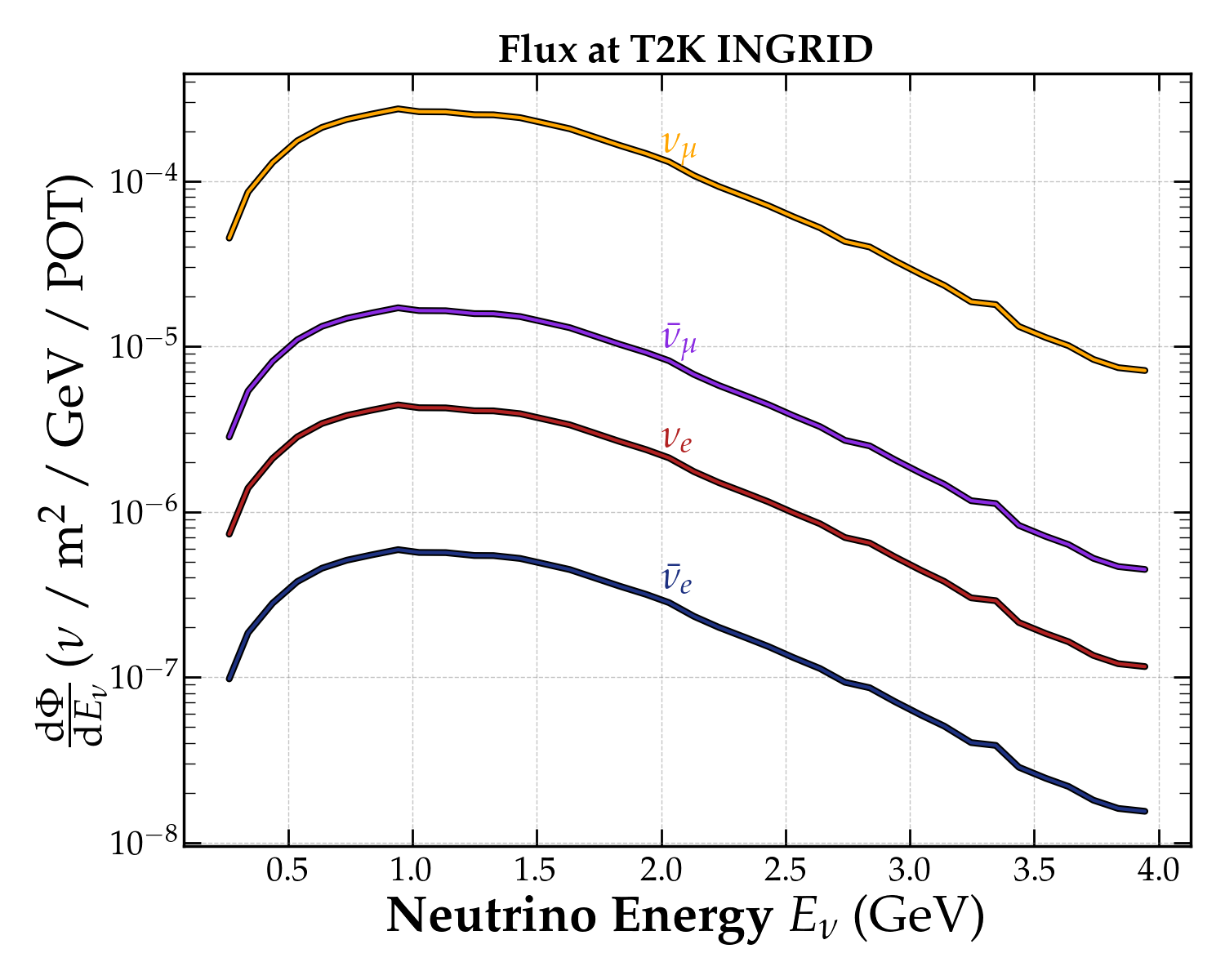}
\includegraphics[width=0.48\textwidth]{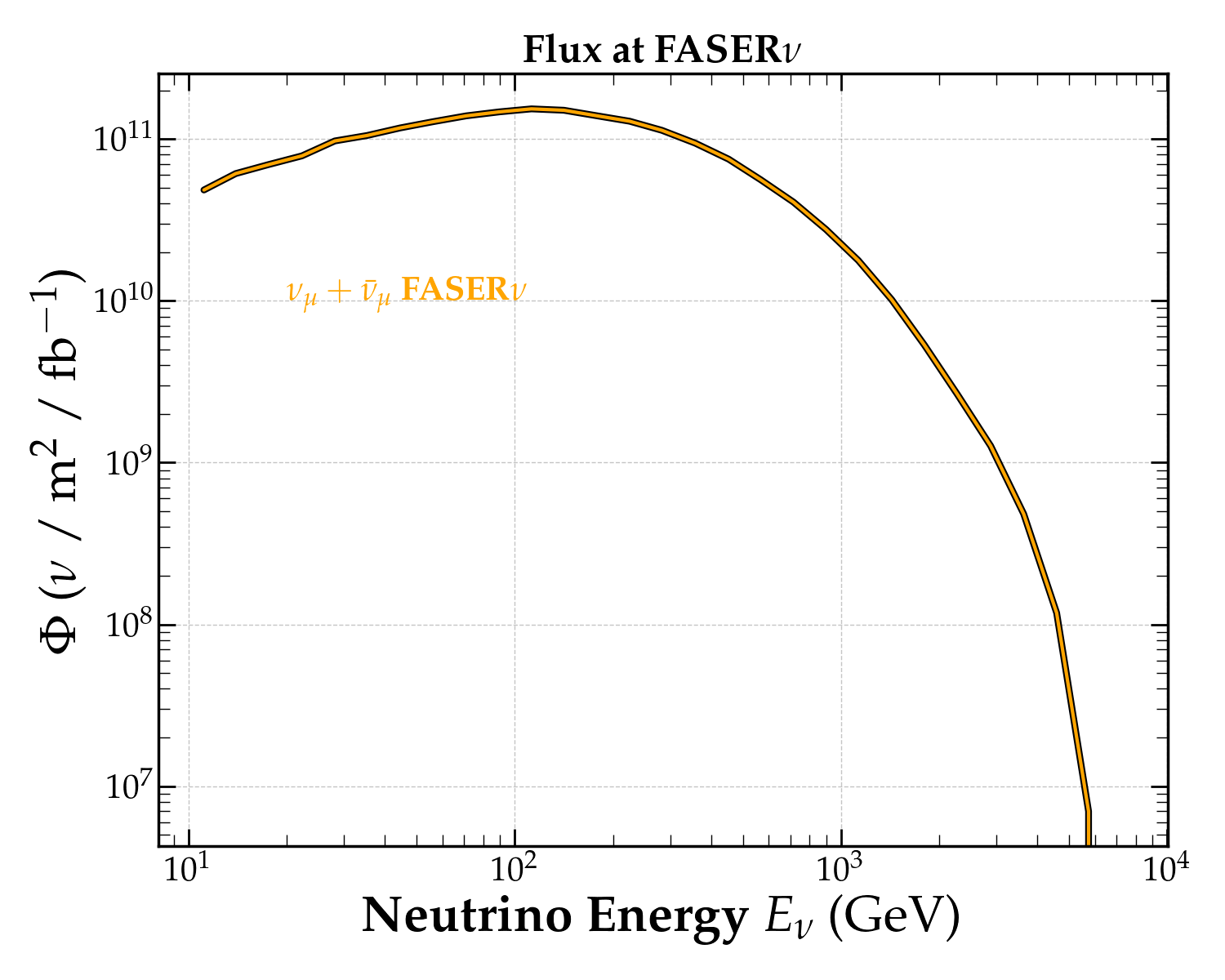}
\caption{Neutrino and antineutrino flux distributions for the accelerator neutrino experiments DUNE, MINOS/MINOS+, T2K considered here. Also shown (bottom right panel) is the collider neutrino flux at FASER$\nu$. See text for details.}
\label{fig:flux}
\end{figure*} 
In Fig.~\ref{fig:flux}, we present the neutrino fluxes at the accelerator neutrino experiments considered here. For DUNE, we show the standard CP-optimized (solid) and tau-optimized (dashed) fluxes for the forward horn current polarity (neutrino) mode on top left panel and antineutrino mode on top right panel from Ref.~\cite{Fields}. Here we consider both electron and muon neutrino (and antineutrino) fluxes at the ND, but the muon neutrino fluxes are orders of magnitude larger.   The MINOS (solid) and MINOS+ (dashed) fluxes for neutrino (middle left) and the MINOS flux for antineutrino (middle right) mode are taken from Refs.~\cite{MINOS:2007ixr, MINERvA:2016iqn}. Note that MINOS+ did not run in antineutrino mode. The T2K INGRID near detector fluxes (bottom left) are taken from Ref.~\cite{T2K:2015cxp}. Note that these fluxes are the same in T2K Phase-1 and Phase-2. We do not consider NO$\nu$A here, because the NO$\nu$A ND neutrino fluxes peak at slightly lower energies than the MINOS flux, $E_\nu^{\rm peak}\approx 2$ GeV~\cite{Maan:2015bzx}, which is below the threshold for tau trident production (see Table~\ref{tab:Thresholds}).

As for collider neutrino experiments, we take FASER$\nu$/FASER$\nu$2 as a prototypical example. The $\nu_\mu + \bar{\nu}_\mu$ fluxes shown here (bottom right) are taken from Ref.~\cite{FASER:2019dxq}.  

\section{Trident Events at Other Accelerator/Collider Neutrino Experiments}
\label{app:exp}
\begin{figure*}[t!]
    \centering
 \includegraphics[scale=0.141]{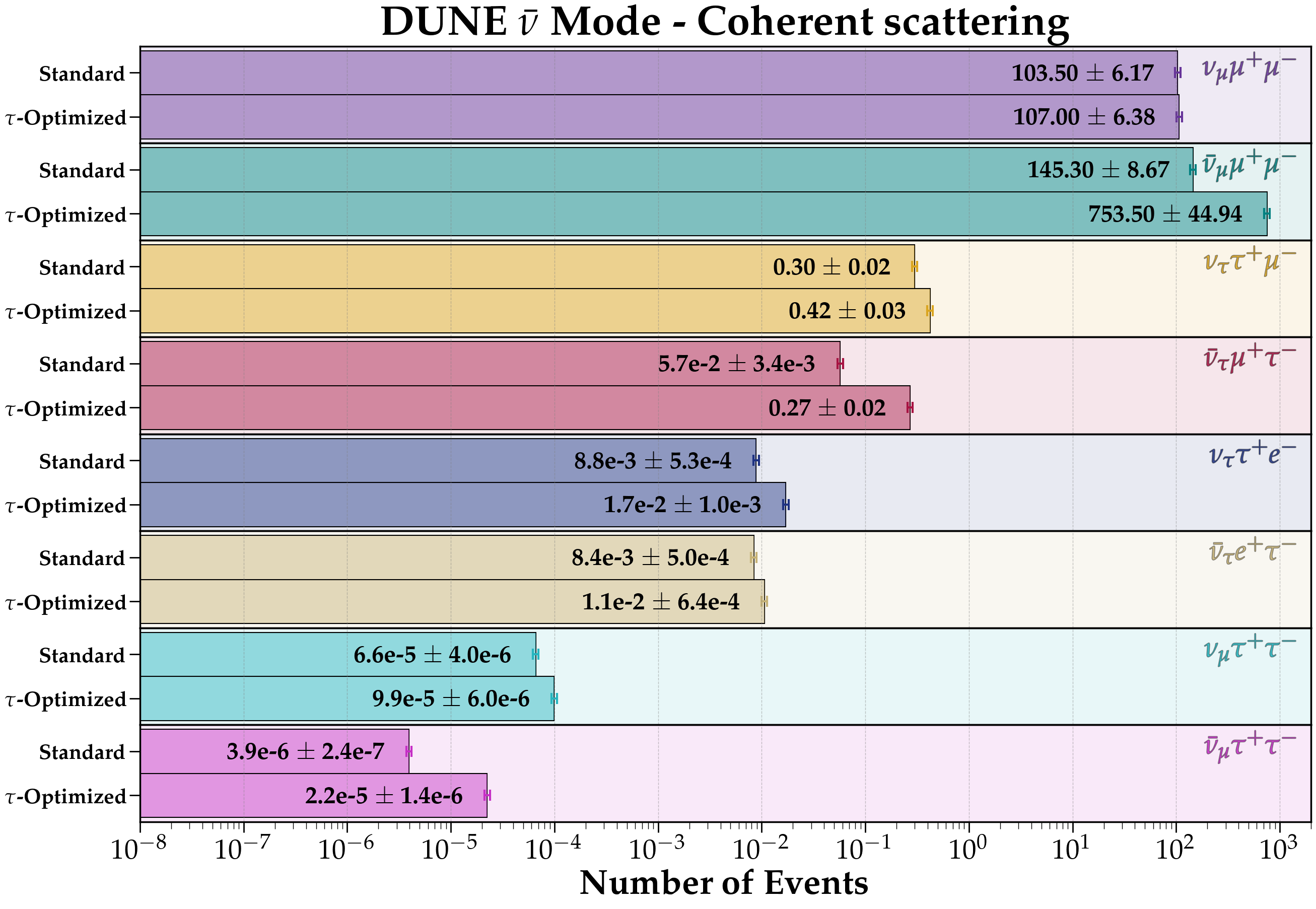}
     \includegraphics[scale=0.141]{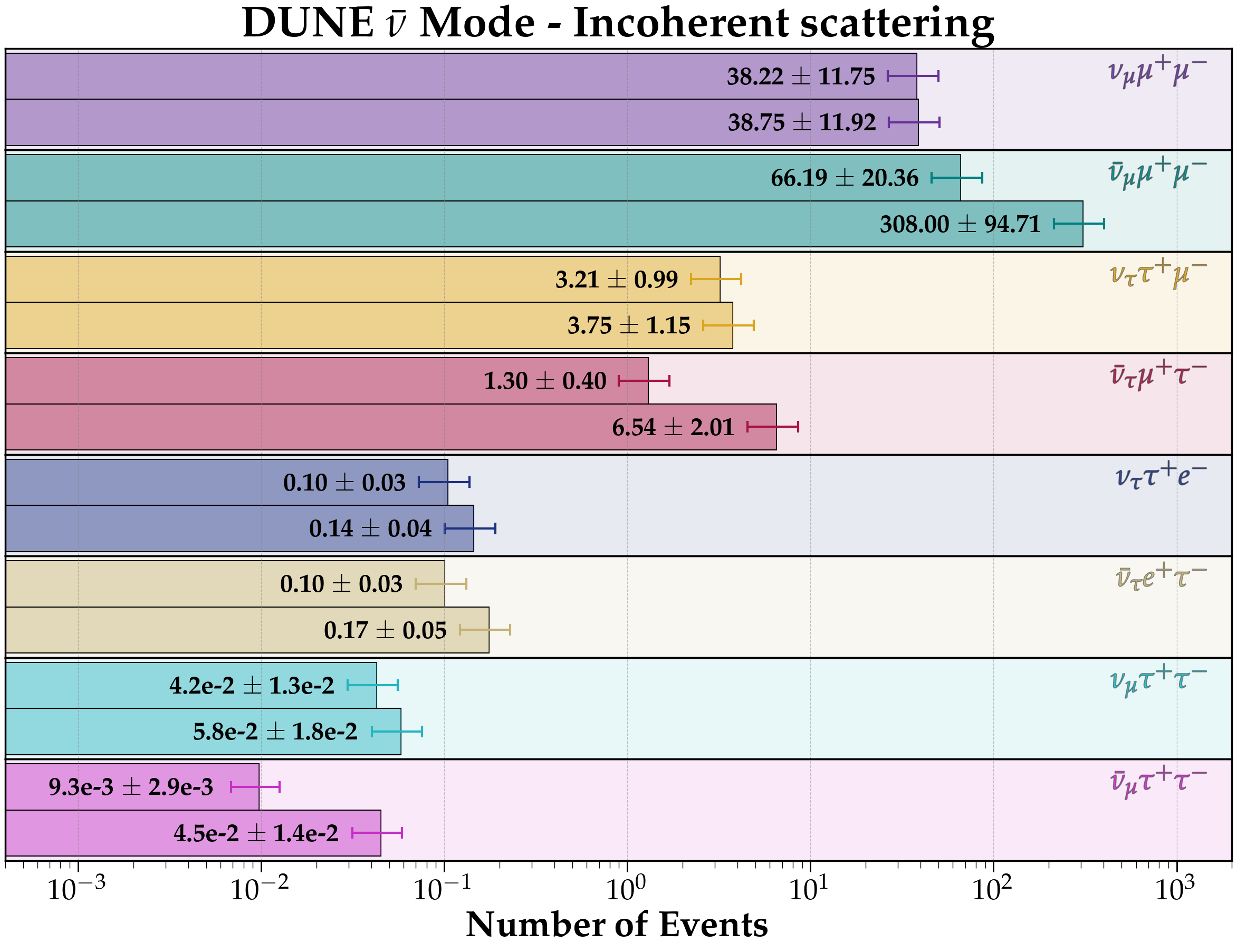}
    \caption{Number of trident events from coherent (left) and incoherent (right) scattering at DUNE near detector with 67 ton of argon and $3.3\times10^{21}$ POT in antineutrino  mode. }
    \label{fig:DUNE-antineutrino-mode}
\end{figure*}

Figure~\ref{fig:DUNE} 
in the main text showed the number of trident events at DUNE ND in the neutrino mode. Figure~\ref{fig:DUNE-antineutrino-mode} shows the number of events in the antineutrino mode with the same 67 tons of detector mass and $3.3\times 10^{21}$ POT. 
The main reason for relegating this figure to the Appendix is that the tau-optimized mode, which gives the best event rate, may not have an antineutrino run at all, let alone for 3 years. Also, the number of trident events in the antineutrino mode are somewhat smaller than in the neutrino mode. This is due to lower fluxes (see Fig.~\ref{fig:flux}), although the cross sections are the same for neutrinos and antineutrinos.    

\begin{figure*}[t!]
    \centering
    \includegraphics[scale=0.19]{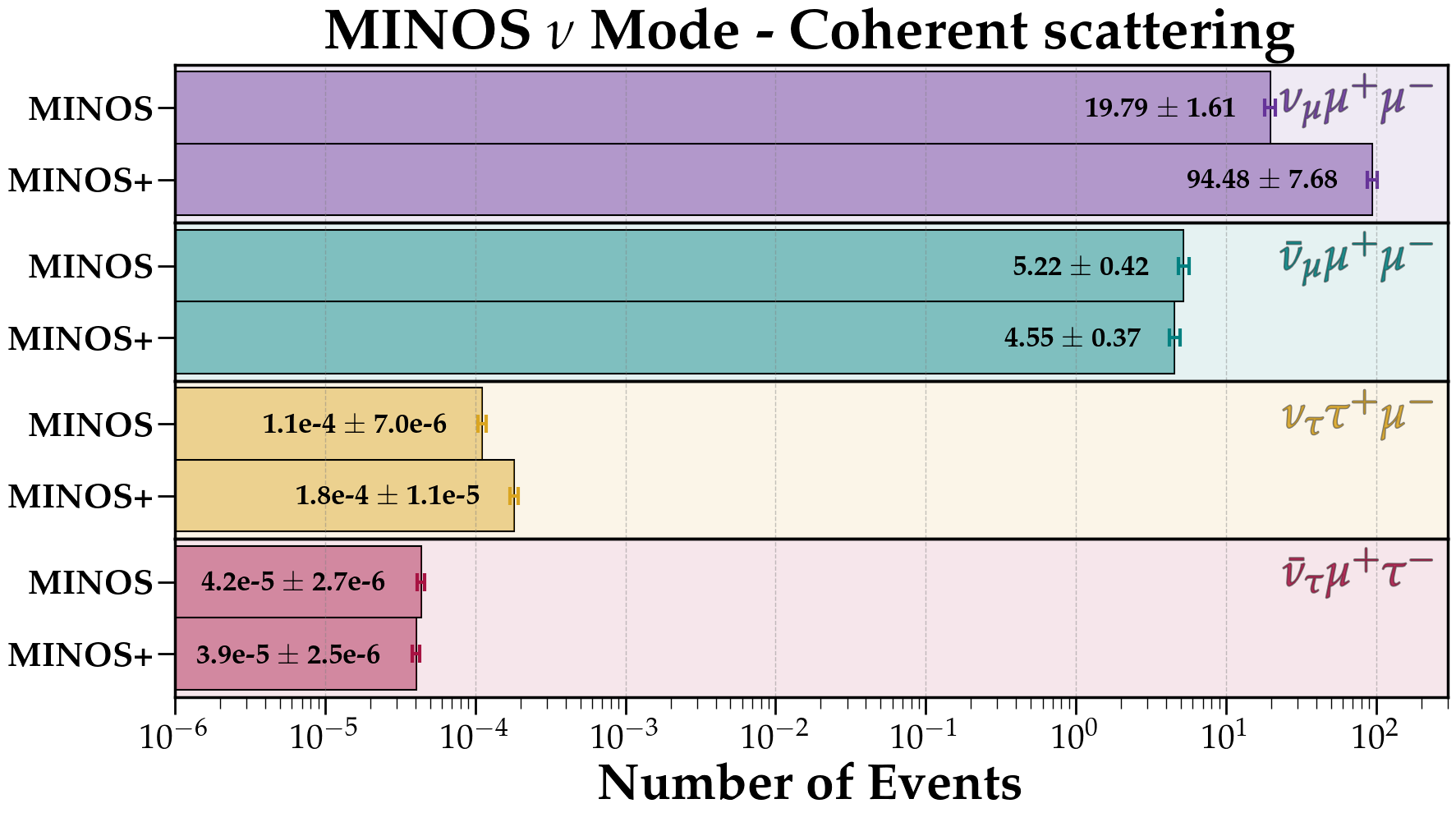}
    \includegraphics[scale=0.19]{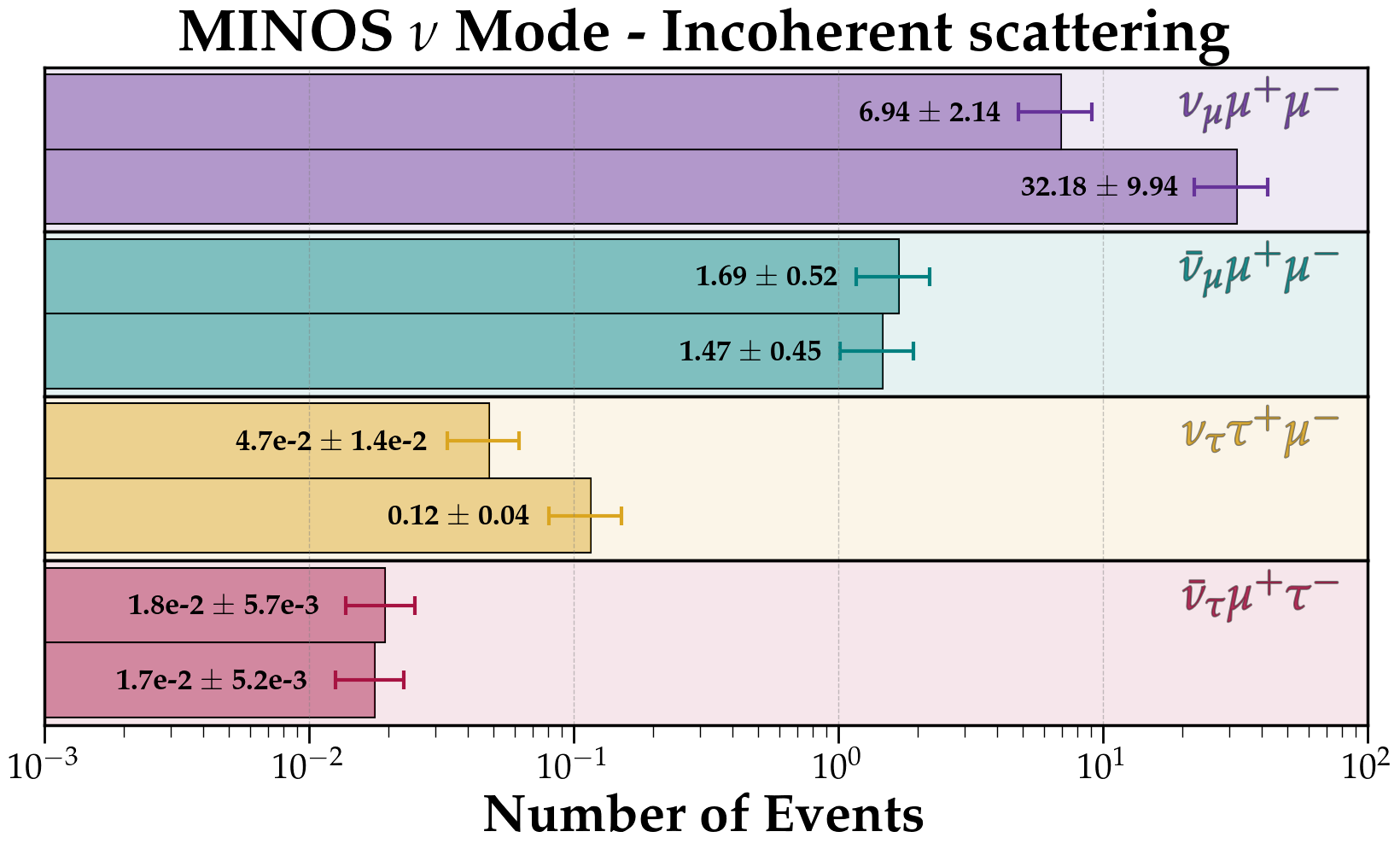}\\
    \includegraphics[scale=0.19]{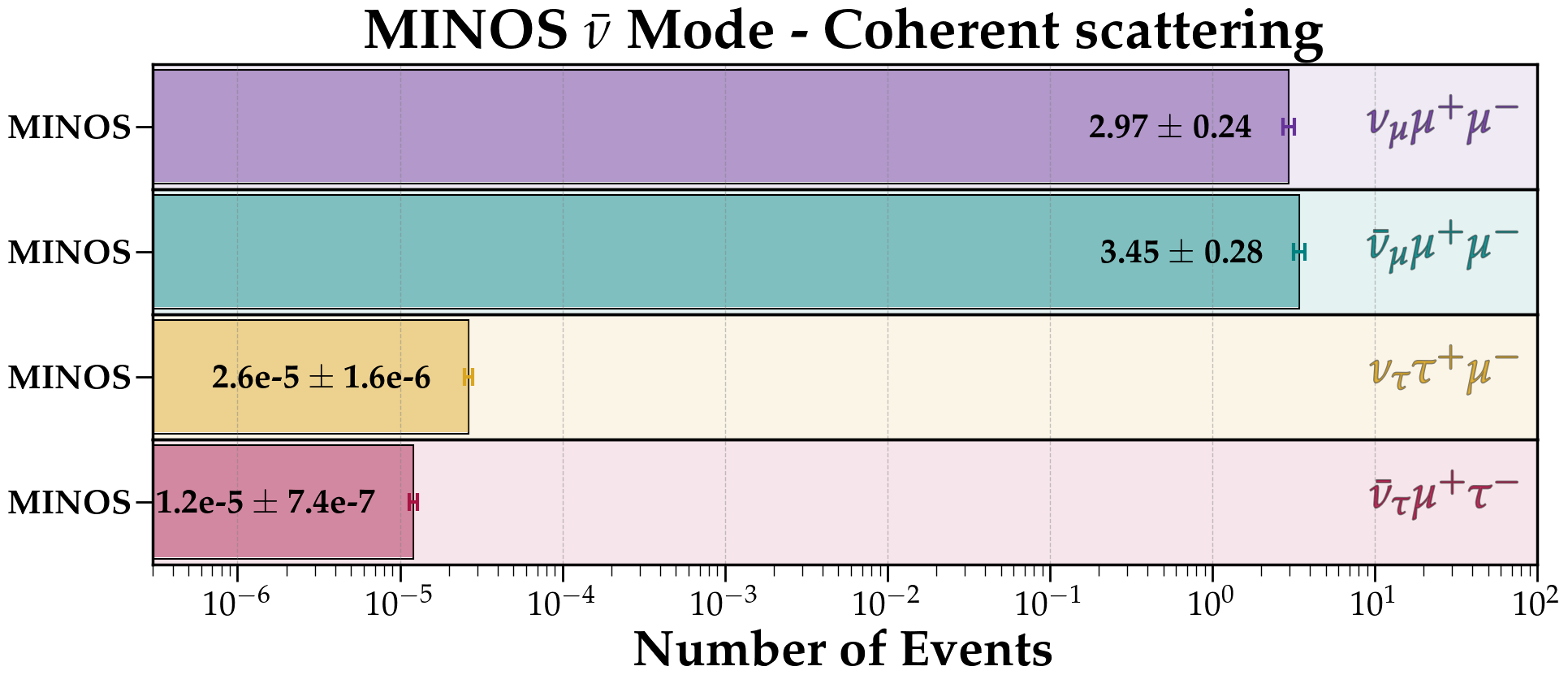}
    \includegraphics[scale=0.19]{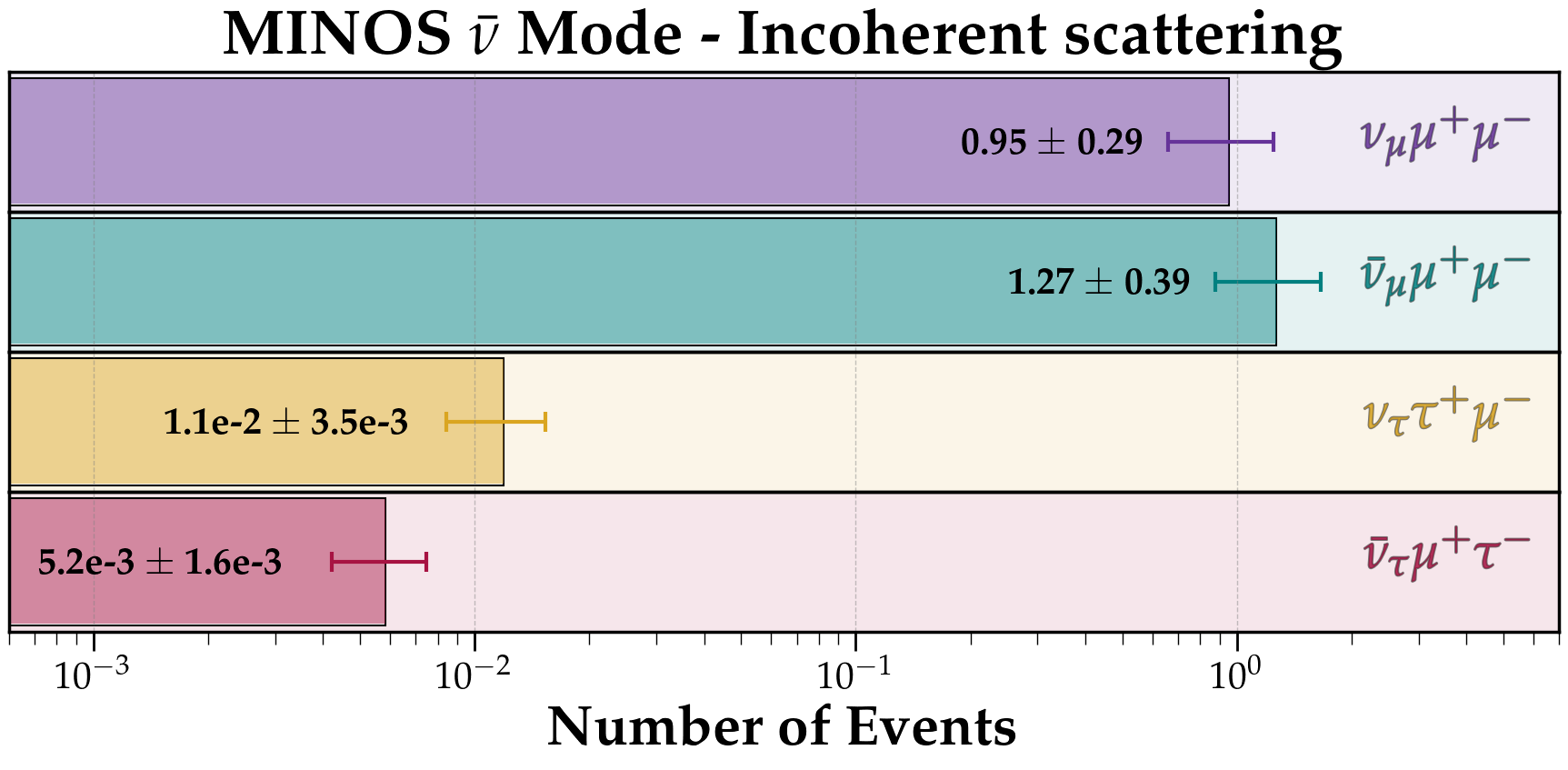}
    \caption{Coherent (left) and incoherent (right) scattering events in the neutrino (top panels) and antineutrino (bottom panels) modes at MINOS and MINOS+ detectors with 28.6 ton iron and $10.56~(9.69)\times 10^{20}$ POT for MINOS (MINOS+) in the neutrino mode and $3.36\times 10^{20}$ POT for MINOS in the antineutrino mode. 
    }
    \label{fig:MINOS_neutrino}
\end{figure*} 

In Figs.~\ref{fig:MINOS_neutrino} and \ref{fig:INGRID}, we show the trident event numbers at MINOS/MINOS+ and T2K INGRID detectors, respectively. The detector material is mostly iron ($^{56}$Fe) in both MINOS/MINOS+ and T2K.
The MINOS ND has a fiducial mass of 28.6 ton and collected $10.56\times 10^{20}$ ($3.36\times 10^{20}$) POT in the neutrino (antineutrino) mode with the low-energy configuration of the NuMI beam having peak neutrino energy $E_\nu^{\rm peak}\approx 3$ GeV~\cite{2018npa..confE.423A}.
Therefore, the number of tau trident events is negligible, as can be seen from Fig.~\ref{fig:MINOS_neutrino}. 
MINOS+ ran with the same detector, but in the medium-energy configuration of NuMI, having $E_\nu^{\rm peak}\approx 7$ GeV and collecting $9.69\times 10^{20}$ POT in the neutrino mode~\cite{2018npa..confE.423A}. 
MINOS+ did not run in antineutrino mode. 
Because of the higher $E_\nu^{\rm peak}$, MINOS+ has slightly larger number of trident events than MINOS, but it is still negligible for the tau tridents due to relatively low POT. 
We do not include MINER$\nu$A here, because although it collected similar POT as MINOS, its fiducial mass is only 8 ton, and therefore, the number of tau tridents will be smaller than the MINOS values. 

\begin{figure*}[t!]
    \centering
\includegraphics[scale=0.168]{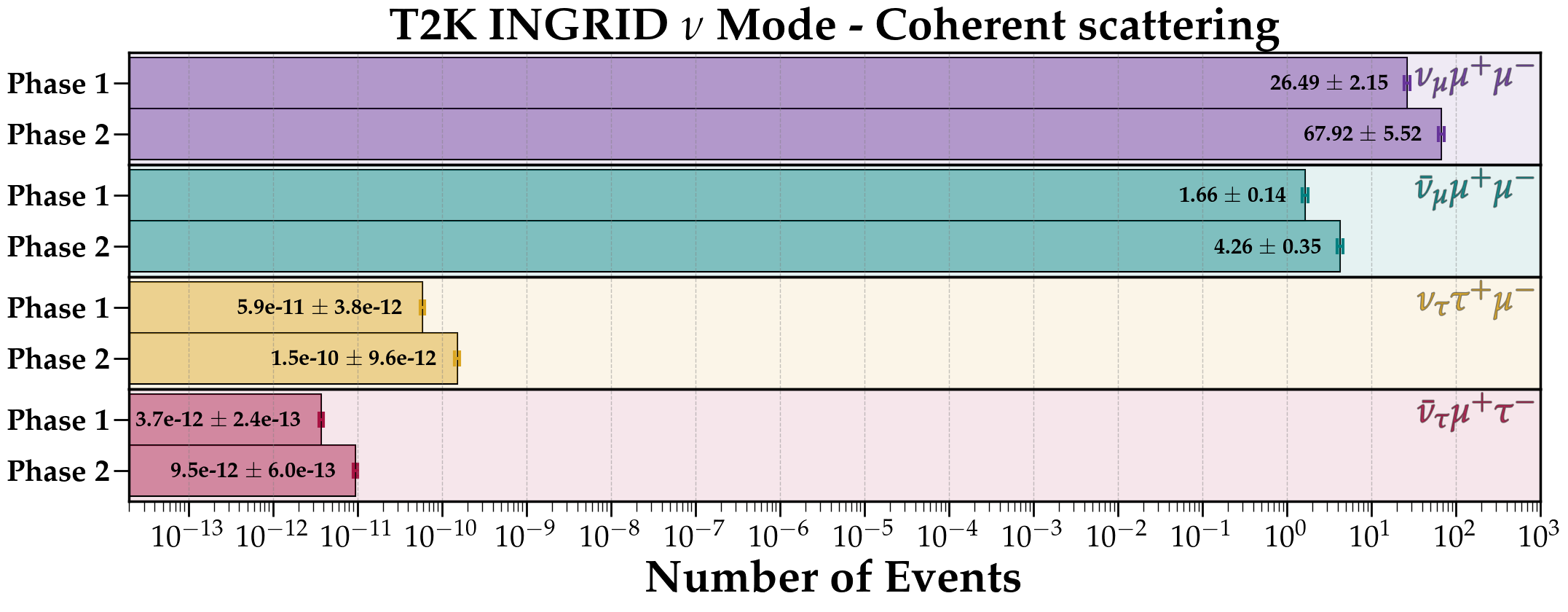}
    \includegraphics[scale=0.168]{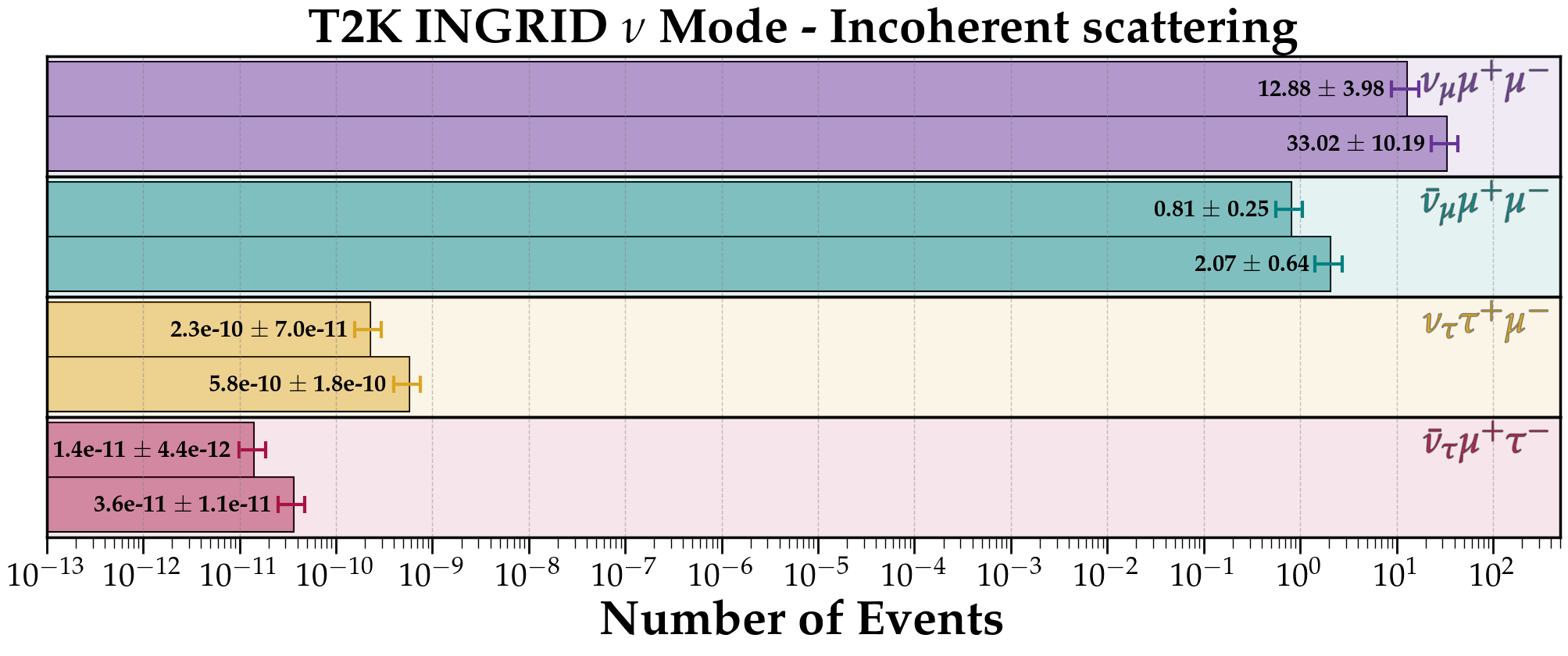}
    \caption{Coherent (left) and incoherent (right) scattering events at T2K INGRID near detector with $99.4$ ton of iron and $3.9~(10)\times 10^{21}$ POT in Phase-1 (2). 
    }
    \label{fig:INGRID}
\end{figure*}

At T2K, the on-axis near detector INGRID has a total fiducial mass of 99.4 t and collected $3.9\times 10^{21}$ POT in each neutrino and antineutrino modes in Phase-1. 
Phase-2 is expected to collect $1.0\times 10^{22}$ POT in each neutrino and antineutrino modes~\cite{T2K:2016siu}. Although both POT and detector mass are larger for T2K, its flux only ranges from 0-4 GeV, and therefore, the number of tau tridents is even smaller than the MINOS case, as shown in Fig.~\ref{fig:INGRID}.  

\begin{figure*}[t!]
    \centering
    \includegraphics[scale=0.2]{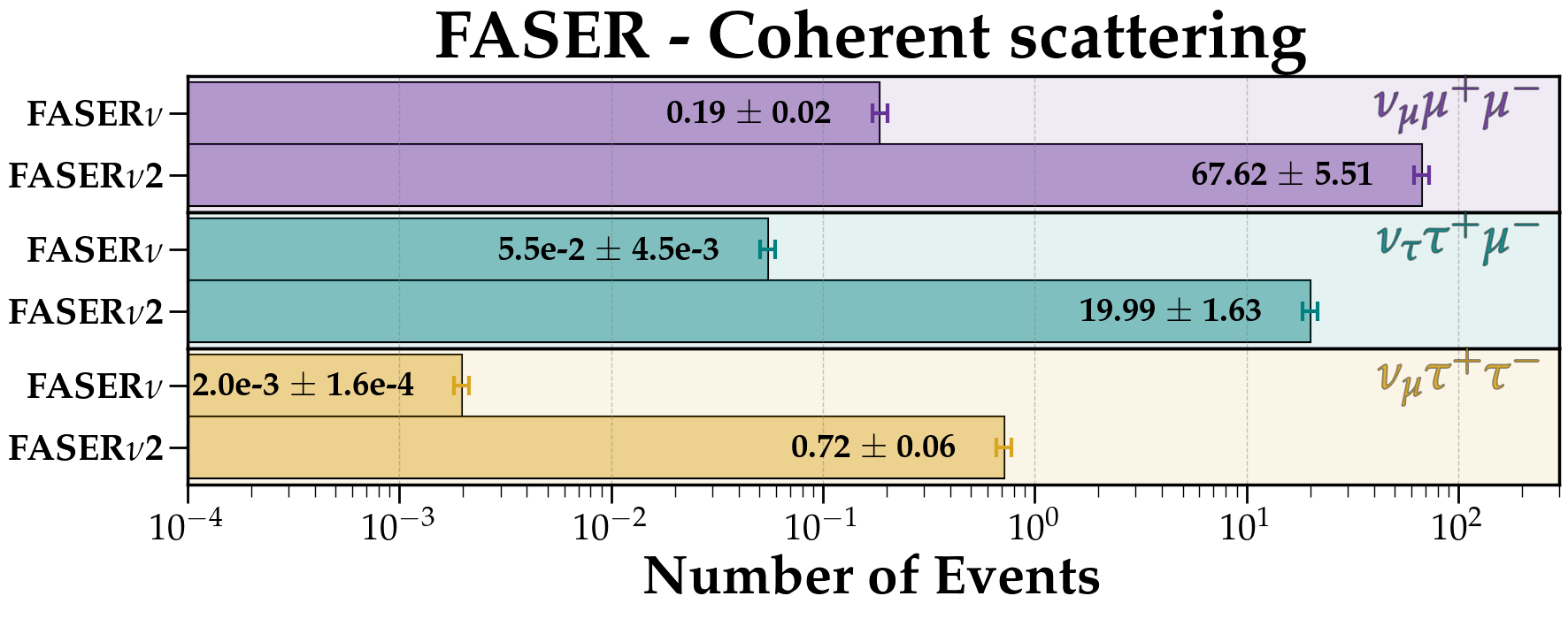}
\includegraphics[scale=0.2]{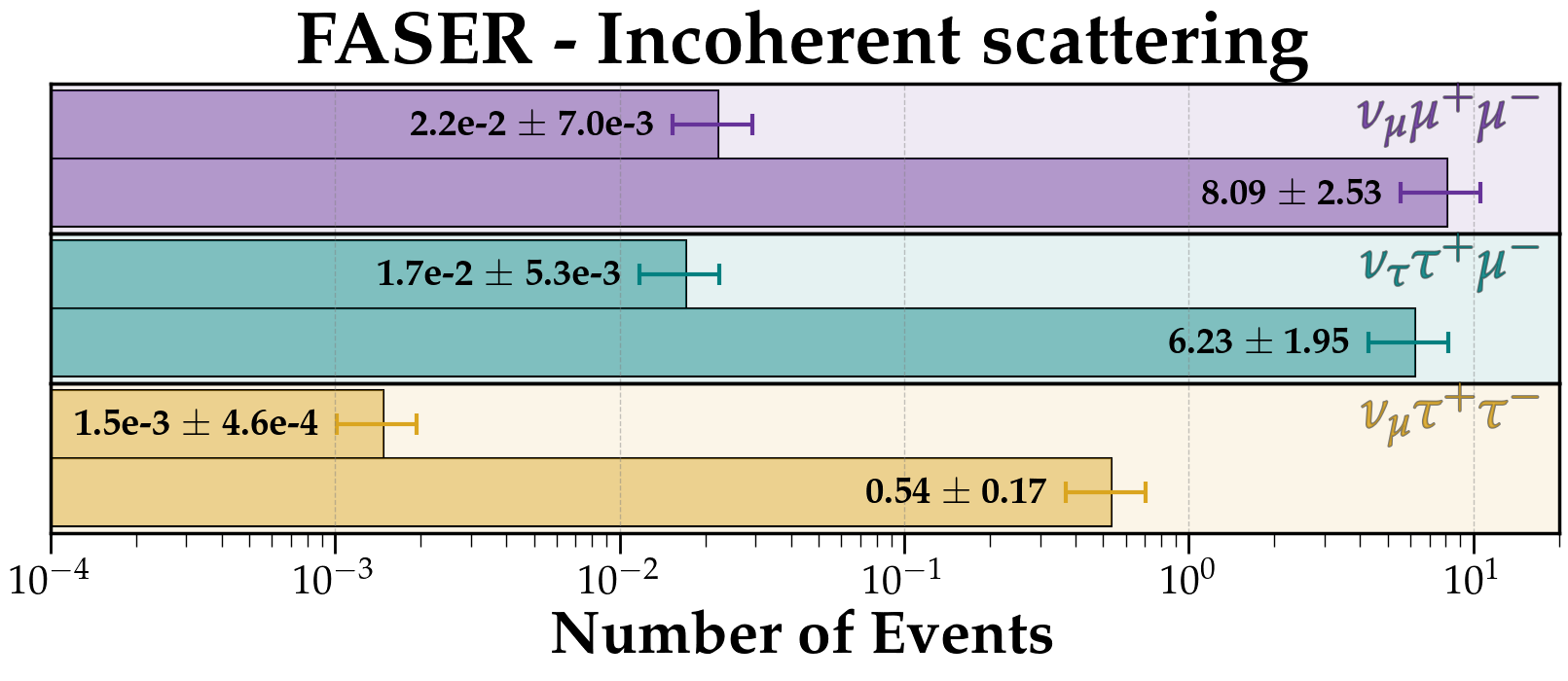}
    \caption{Coherent (left) and incoherent (right) scattering events at FASER$\nu$ and FASER$\nu$2 detectors with fiducial mass of 1.1 ton and 20 ton tungsten, and integrated luminosity of 150 fb$^{-1}$ and 3 ab$^{-1}$, respectively.}
    \label{fig:FASER}
\end{figure*}

We also consider collider neutrinos at FASER$\nu$/FASER$\nu$2 and calculate the number of trident events, as shown in Fig.~\ref{fig:FASER}. 
These numbers were not available for any of the trident channels until very recently~\cite{Francener:2024wul, Altmannshofer:2024hqd}. 
While Ref.~\cite{Francener:2024wul} has no mention of tau tridents, Ref.~\cite{Altmannshofer:2024hqd} estimates those numbers to be small and experimentally challenging to probe. 
Our numbers corroborate this observation.   
Note that the neutrinos reaching FASER$\nu$ are of much higher energy than the accelerator neutrinos, in the range of 10 GeV--1 TeV or so. 
Therefore, the coherent scattering will give the dominant contribution to the trident processes (see Fig.~2)
at FASER$\nu$/FASER$\nu$2. 
The FASER$\nu$ detector is made of tungsten ($^{184}$W) and has a fiducial mass of 1.1 ton. 
FASER$\nu$ is expected to collect 150 fb$^{-1}$ integrated luminosity during the current Run 3 of LHC. 
The proposed FASER$\nu$2 detector~\cite{Feng:2022inv} will have a mass of 20 ton and is expected to collect 3 ab$^{-1}$ integrated luminosity at HL-LHC.  
In Fig.~\ref{fig:FASER}, we show the expected number of trident events at FASER$\nu$/FASER$\nu$2 with these configurations. Since the neutrino luminosities are readily available for FASER$\nu$/FASER$\nu$2, we utilize a similar method as Eq.~(5) 
but with the measured neutrino luminosity rather than $N_\text{POT}$ to estimate the number of  events. 
We find that while FASER$\nu$ will not have enough statistics to observe tau tridents, FASER$\nu$2 will get about 20 events from coherent scattering and 6 from incoherent scattering. 
Although these numbers are similar to what we got for DUNE, there is a potentially large intrinsic $\nu_\tau$ background of ${\cal O}(2000)$ events~\cite{Bai:2021ira,Bai:2022jcs, reno} from decays of charmed mesons, which are copiously produced at the LHC energy.  
Therefore, we foresee that distinguishing ${\cal O}(20)$ tau trident events from this large $\nu_\tau$-induced CC background will be quite challenging.  In fact, this is going to be a general problem for other future collider neutrino experiments as well, such as  SND@LHC~\cite{SNDLHC:2022ihg}, SND@SHiP~\cite{SHiP:2015vad} and FLArE~\cite{Batell:2021blf}.

\section{Tau Trident Event Distributions at DUNE} \label{app:dist}
\begin{figure*}[t!]
    \centering
    \includegraphics[width=0.49\textwidth]{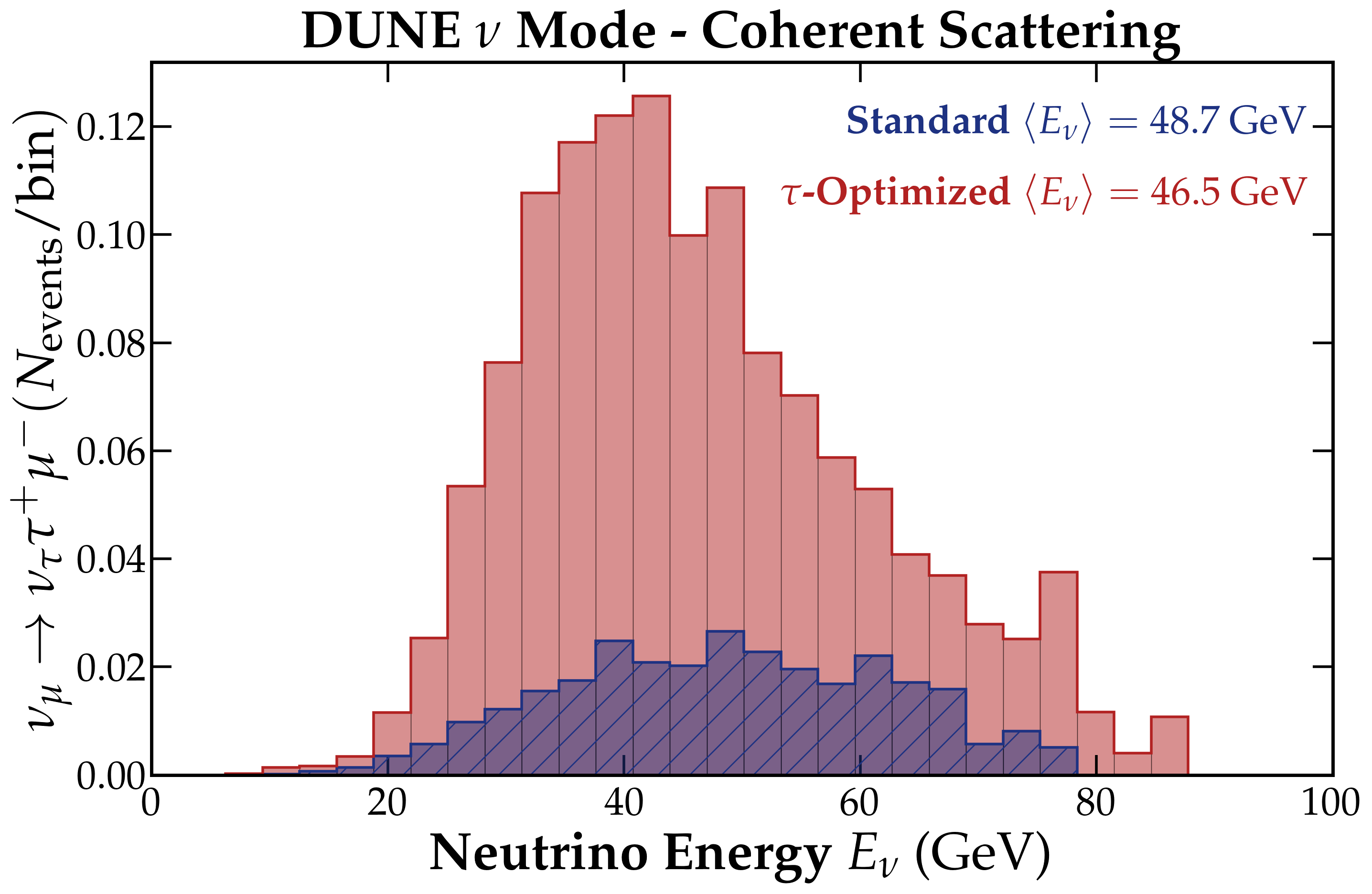}
    \includegraphics[width=0.49\textwidth]{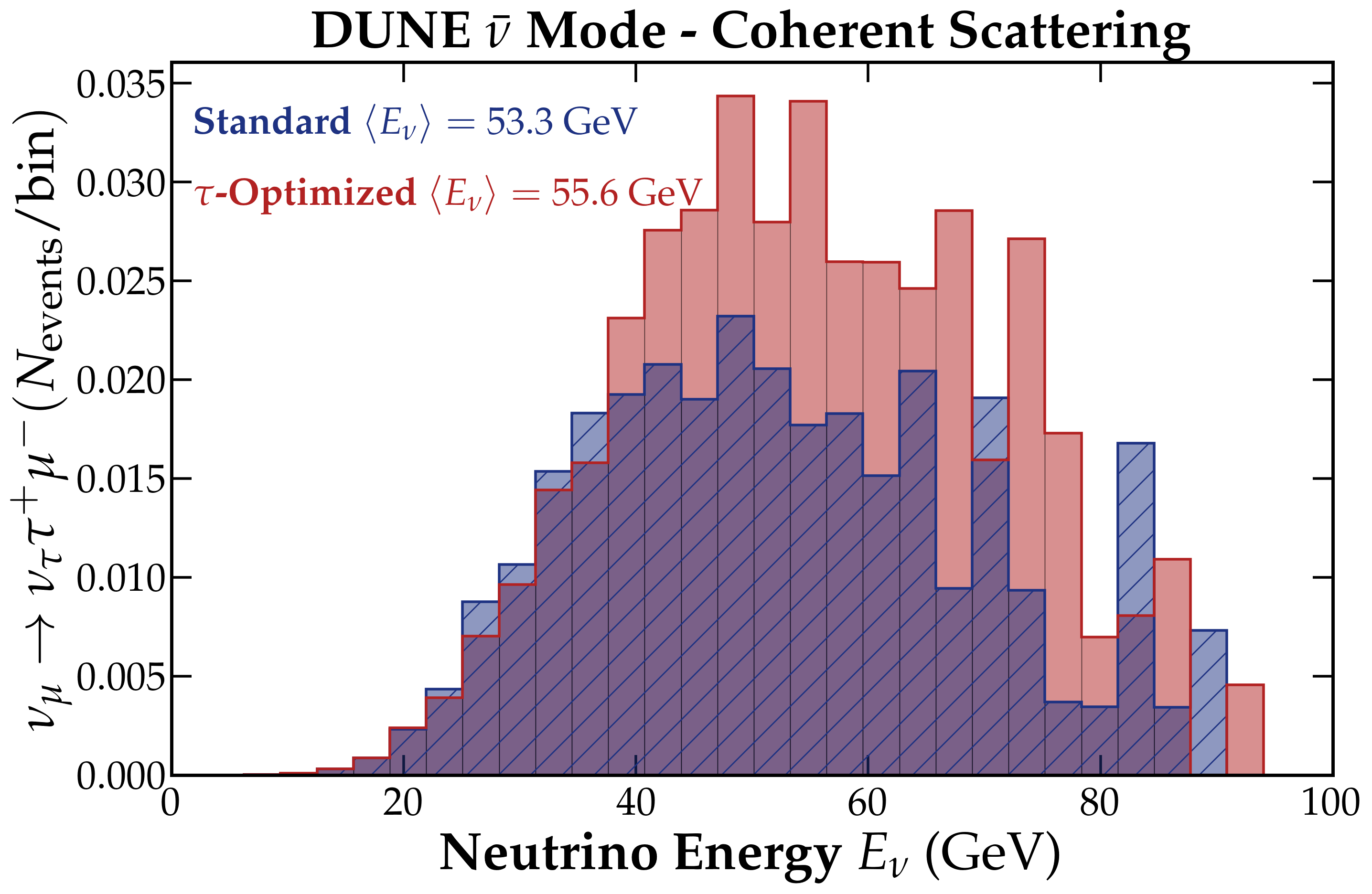}\\
     \includegraphics[width=0.49\textwidth]{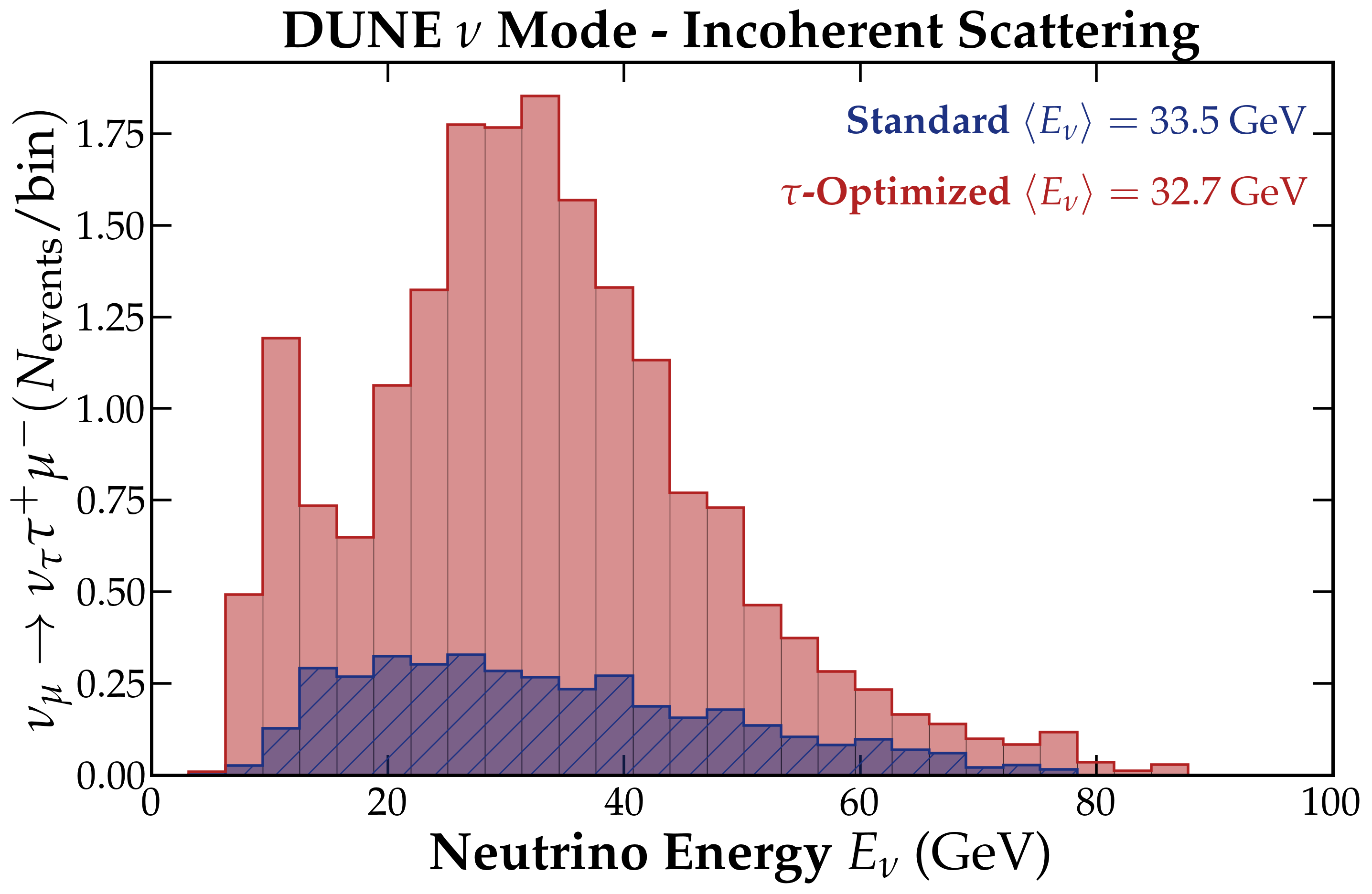}
    \includegraphics[width=0.49\textwidth]{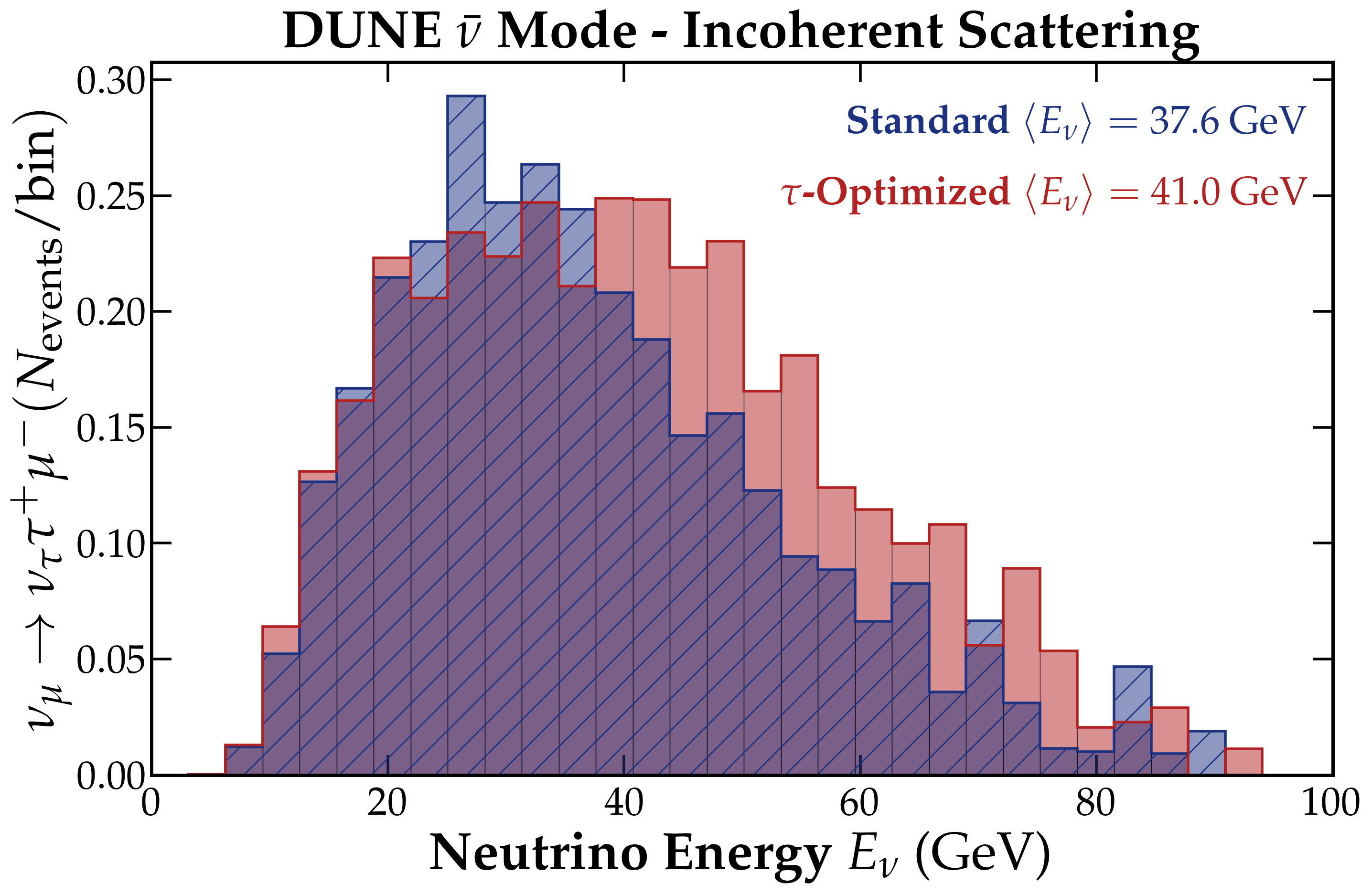}
    \caption{Number of events per bin as a function of neutrino energy $E_\nu$ for the single-tau trident $\nu_\mu N \to \nu_\tau N \tau^+ \mu^-$ with coherent (top) and incoherent (bottom) scatterings in DUNE neutrino (left) and antineutrino (right) beam mode for the standard (blue) and tau-optimized (red) horn configurations. We have taken 67 ton fiducial mass ND and an exposure of $3.3\times 10^{21}$ POT.}
    \label{fig:event_dist}
\end{figure*} 

In Fig.~\ref{fig:event_dist}, we show the number of events per bin as a function of neutrino energy $E_\nu$ for the single-tau trident $\nu_\mu N \to \nu_\tau N \tau^+ \mu^-$ with coherent (top) and incoherent (bottom) scatterings in DUNE neutrino (left) and antineutrino (right) beam mode for the standard (blue) and $\nu_\tau$-optimized (red) horn configurations. The average neutrino energy for each distribution is shown for reference, which is used for the missing momentum distribution in Fig.~\ref{fig:RMiss_1tau}. 
Note that the peaks of these distributions occur at energies much higher than those at which the DUNE fluxes peak. This is simply because the trident processes involving tau leptons in the final state are phase-space suppressed and require a higher energy threshold. This is the main reason why the number of tau tridents is much smaller, as compared to the muon tridents.

\bibliographystyle{utphys}
\bibliography{main_bibl}
\end{document}